\documentclass[journal]{IEEEtran}%
\ifCLASSINFOpdf
\else
\fi
\usepackage{graphicx}
\usepackage{makecell}
\graphicspath{ {./png} }
\usepackage{amsmath}
\usepackage{algorithm}
\usepackage{algpseudocode}
\usepackage{subfigure}
\usepackage{amsfonts}
\usepackage{cite}
\usepackage{xcolor}
\usepackage{soul}

\begin{document}

\title{Modeling, Design, and Verification \\ 
of An Active Transmissive RIS}

\author{Rongguang~Song,
        Haifan~Yin,~\IEEEmembership{Senior Member,~IEEE,}
        Zipeng~Wang,
        Taorui~Yang,
        and~Xue~Ren,~\IEEEmembership{Member,~IEEE}
\thanks{This work was supported by the Fundamental Research Funds for the Central Universities and the National Natural Science Foundation of China under Grant 62071191.}
\thanks{The corresponding author is Haifan Yin.}
\thanks{R. Song, H. Yin, Z. Wang, and T. Yang are with the School of Electronic Information and Communications, Huazhong University of Science and Technology, Wuhan 430074, China (e-mail: song\_rg@hust.edu.cn; yin@hust.edu.cn; zpw@hust.edu.cn; try@hust.edu.cn).}
\thanks{X. Ren is with the State Key Laboratory of Radio Frequency Heterogeneous Integration, Shenzhen University, Shenzhen, China, and the College of Electronics and Information Engineering, Shenzhen University, Shenzhen, China,  and the Institute of Microelectronics, Shenzhen University, Shenzhen 518060, China. (e-mail: eerenxue@szu.edu.cn).}
}

\maketitle

\begin{abstract}
Reconfigurable Intelligent Surface (RIS) is a promising technology that may effectively improve the quality of signals in wireless communications. In practice, however, the ``double fading'' effect undermines the application of RIS and constitutes a significant challenge to its commercialization. To address this problem, we present a novel 2-bit programmable amplifying transmissive RIS with a power amplification function to enhance the transmission of electromagnetic signals. The transmissive function is achieved through a pair of radiation patches located on the upper and lower surfaces, respectively, while a microstrip line connects two patches. A power amplifier, SP4T switch, and directional coupler provide signal amplification and a 2-bit phase shift. To characterize the signal enhancement of active transmissive RIS, we propose a dual radar cross section (RCS)-based path loss model to predict the power of the received signal for active transmissive RIS-aided wireless communication systems. 
 Simulation and experimental results verify the reliability of the RIS design, and the proposed path loss model is validated by measurements. Compared with the traditional passive RIS, the signal power gain in this design achieves 11.9 dB. 
\end{abstract}


%
\IEEEpeerreviewmaketitle

\section{Introduction}
%
%
%
%
\IEEEPARstart{W}{ith} the rapid advancement of wireless communication technologies in recent years, the deployment of fifth-generation (5G) communication systems has been progressively realized in various countries and regions globally. Concurrently, there is an increasing focus on developing the next generation, namely the sixth-generation (6G) communication systems. Users benefit from enhanced transmission capacity, improved reliability, and reduced latency. However, for researchers in the field of wireless communications, new challenges have emerged with these advanced systems, including cost reduction, energy efficiency, extended coverage, and the increasing number of base stations. Reconfigurable Intelligent Surfaces (RIS), composed of numerous sub-wavelength units, have emerged as an effective solution to these challenges. The unique ability to manipulate the amplitude and phase of electromagnetic (EM) waves in specific ways has been demonstrated as a viable approach to address these challenges \cite{pendry2000negative,pendry2006controlling,zheludev2012metamaterials,shu2023amplifying,basar2019wireless,tang2019subject}. Unlike traditional methods of improving communication quality by optimizing transmitters and receivers, the introduction of RIS enables a higher degree of control and optimization within the transmission path of wireless communication systems. Intelligent manipulation of EM waves by RISs has made the controllability, programmability, and optimization of propagation channels feasible \cite{wu2019towards,kafesaki2012optically}. Furthermore, compared to the previously widely used phased array antennas, which are associated with higher deployment costs, RISs have an obvious advantage in terms of deployment cost and complexity, as RISs do not necessitate additional wired signal sources \cite{jin2020wideband,wen2015wide,arand2017design}. 

In 2011, Yu et al. proposed the concept of the generalized Snell's law, which achieves anomalous refraction of light by incorporating a phase gradient at the interface\cite{Yu2011LightgeneralizedSnell's}. In 2014, the concept of utilizing metamaterial technology for EM wave control through the development of coding, digital, and programmable intelligent metasurfaces was introduced\cite{Kaina2014Shaping,cui2014coding}, providing crucial theoretical and technical support for complex wave manipulation in practical applications.
The working principle of RISs is rooted in array signal processing and reflective optics\cite{Yu2011LightgeneralizedSnell's}, consisting of numerous small units, each containing one or more tunable impedance elements \cite{holloway2012overview}, such as varactor diodes \cite{pei2021ris,sievenpiper2003two,araghi2022reconfigurable} and positive-intrinsic negative diode (PIN diodes) \cite{cao20231,dai2020reconfigurable,hu2021design}. Generally, each unit in a RIS receives multi-channel control signals through the integration of field programmable gate arrays (FPGA) or other controllers, adjusting the interference and reflection characteristics of the incident waves, thus enabling dynamic beamforming \cite{wan2016field,shcherbakov2017ultrafast,yang20161,yang2016programmable,li2016reconfigurable}. Therefore, inserting an RIS between the transmitter and receiver can effectively mitigate EM wave signal interference, significantly enhancing the signal-to-noise ratio, signal coverage, and data transmission rate \cite{basar2019wireless}.

Recently, numerous passive RIS structure designs have been proposed and researched. 
However, most existing designs are reflective RISs. For instance, the authors of \cite{cao20231} present the design and implementation of a 1-bit time-modulated reflectarray (TMRA), where PIN diodes act as switches in each unit,
 the RIS successfully achieves beam scanning within a range of 0$^{\circ}$-50$^{\circ}$ and demonstrates ultra-low side lobe levels (SLL) in the scattering pattern.
In \cite{sievenpiper2003two}, the varactor diodes are tuned by bias voltages to achieve desired phase shifts, enabling the 1100 controllable elements RIS operating at 5.8 GHz to reflect incident waves as a beam pointing at the receiver, with a 27 dB power gain observed in short-distance outdoor measurement. These designs 
indicate that, in specific environments, RIS 
improve the cell-edge coverage. Considering the requirement for RIS to be versatile in various communication environments, the field is experiencing a growing consideration for the design of transmissive RIS, complementing the reflective RIS types previously discussed. In \cite{li2016transmission}, a design is proposed featuring a programmable unit cell with PIN diodes, enabling a 2-bit phase shift and utilized for imaging measurement in a 10 $\times$ 10 array. Another study in \cite{tang2022transmissive} introduces a 16 $\times$ 16 element transmissive RIS prototype, achieving a gain of 22.0 dBi at 27 GHz and demonstrating two-dimensional beamforming with a scanning angle of ±60 degrees. The proposal of transmissive RIS designs effectively complements reflective RIS, providing more application scenarios for RIS-aided systems.

With the continued development of RIS research, issues such as transmission losses introduced by RIS have become increasingly evident. Several studies have shown that in RIS-aided systems, wireless signals propagating from the transmitter to the receiver incur path losses in two segments: from the transmitter to the RIS and from the RIS to the receiver\cite{zhang2022active,dong2021active,shu2023amplifying,rao2023active,wu2022wideband}. This results in a ``double fading" effect that significantly increases the overall path loss of the RIS-aided system\cite{zhang2022active}, thereby restricting the applicability of RIS in various communication scenarios. Additionally, transmission through transmissive RIS can incur insertion losses, which, in some cases, may be as high as 5.7 dB or even more \cite{yang2016study}. Losses due to phase shifts in electronic components also cannot be ignored. These losses significantly limit the propagation distance in RIS-aided systems. In \cite{wu2019beamforming}, the authors analyzed the phase quantization effect caused by the number of phase shifts. 
The results show that phase quantization errors associated with discrete phase shifts inevitably introduce performance degradation of RIS. Compared to 1-bit reconfigurable reflect array antennas, 2-bit designs have superior SLL, achieving lower losses while ensuring low complexity. With these shortcomings identified, active RIS with signal amplification capabilities has emerged as a new topic to improve signal quality further.The work \cite{rao2023active} proposes an active RIS, achieving 2-bit adjustment and amplification of signals through power amplifier chips and four single-pole double-throw (SPDT) switches. Experimental results demonstrate that the designed 2×2 active RIS prototype can enhance the received power by approximately 8.5 dB compared to passive RIS. Furthermore, in \cite{wu2022wideband}, the authors introduce a wideband amplifying reconfigurable intelligent surface (ARIS), which enhances reflected EM signals by integrating power amplifiers (PAs). 
The paper details a dual orthogonal hourglass-slot aperture-coupled patch element and an integrated power synthesis and distribution network, achieving a gain of 7.7 to 12.2 dB over a broad operational bandwidth of 5.0 to 6.0 GHz, thereby reducing the number of required power amplifiers and associated system costs. In \cite{Che2021Dual-PolarizedMetasurface}, the authors present a dual-polarized nonreciprocal active metasurface (SAA-MTS) that achieves broadband amplification from 4.88 to 5.19 GHz with a peak gain of 10.4 dB.

Developing a comprehensive model of an active transmissive RIS-aided system is imperative to advance the practical application of transmissive RIS technology in communication systems. Existing research on RIS models primarily hinges on Huygens's principle and antenna theory\cite{Diebold2023Reflect,li2016transmission,tang2022transmissive,li2023towards}, providing foundational theoretical support. According to the standard Huygens's principle, the scattered electric field emerged from the transmission-type metasurface \cite{li2016transmission}. In \cite{tang2022transmissive}, a free space channel model based on the antenna theory for the transmissive RIS-aided communication system is proposed. This model is analogous to those used for reflective RISs and incorporates beamforming techniques for practical RISs with discrete phase shifts.
In \cite{Diebold2023Reflect}, the authors propose a patch reflectarray modeling approach by describing each patch as a pair of polarizable magnetic dipoles. The method extracts effective polarizabilities through full-wave simulations and applies them to the design of variable patch size and binary state reflectarrays, accurately predicting beam patterns consistent with full-wave simulation results.
In \cite{li2023towards}, the authors present the modeling of a transmissive RIS transceiver-enabled uplink multi-user communication system, taking both far-field and near-field into account. Active transmissive RIS offers significant advantages in resolving non-line-of-sight (NLOS) communication challenges by actively adjusting signals through integrated amplifiers. Most existing research on RIS is focused on conventional passive ones. At the same time, the lack of comprehensive modeling studies on active transmissive RIS, specifically regarding its radiation patterns and signal transmission efficiency, impedes its effective incorporation into practical communication systems.


Considering the existing research gaps in RIS design, this paper proposes a novel transmissive RIS for the 2.6 GHz frequency band. This design incorporates active electronic components to amplify the power and is capable of a 2-bit phase shift. 
Our contributions are as follows:

\begin{itemize}
\item A novel 2-bit active transmissive RIS element: We develop a 2-bit programmable amplifying transmissive RIS, utilizing a quadrifilar directional coupler and SP4T switch, augmenting traditional RIS capabilities with integrated power amplification and 2-bit phase shift. The RIS operates at 2.6 GHz, which is a typical 5G frequency band of China Mobile. Based on this foundation, we have also proposed a full-duplex active transmissive RIS design.
    
\item A dual radar cross section (RCS)-based model for active transmissive RIS-aided wireless communication: We propose a model to quantify the path loss in active transmissive RIS-aided wireless communication systems. This model incorporates dual RCS  at both the receiving and transmitting ends of the transmissive RIS to assess path loss accurately. Measurements validate this model.

\item Verification of a $4 \times 8$ active transmissive RIS array: This paper details the fabrication and validation of a 32-element RIS array. The array demonstrates notable capabilities in 2-bit phase shift beamforming and adaptive power control, achieving up to 12 \text{dB} power gain. This empirical evaluation substantiates the practical applicability of the proposed active RIS design.
\end{itemize}

The rest of this paper is organized as follows. Section II introduces the modeling of the transmissive RIS-aided wireless communication system. In Section III, an active transmissive RIS unit structure is proposed. Section IV shows a prototype of the transmissive RIS, and measurements of the structural design and system model are conducted.

\section{System Model}
This section presents a novel path loss model to delineate the path loss characteristics in an active transmissive RIS-aided wireless communication system.

\subsection{Received Signal Model}
In a three-dimensional (3D) space, we consider a transmissive RIS-aided wireless communication system that consists of two horn antennas, serving as the transmitter (TX) and the receiver (RX), respectively, and a uniform planner array (UPA) of active transmissive RIS elements.

\begin{figure}[!htbp]
	\centering
	\includegraphics[width=0.4\textwidth]{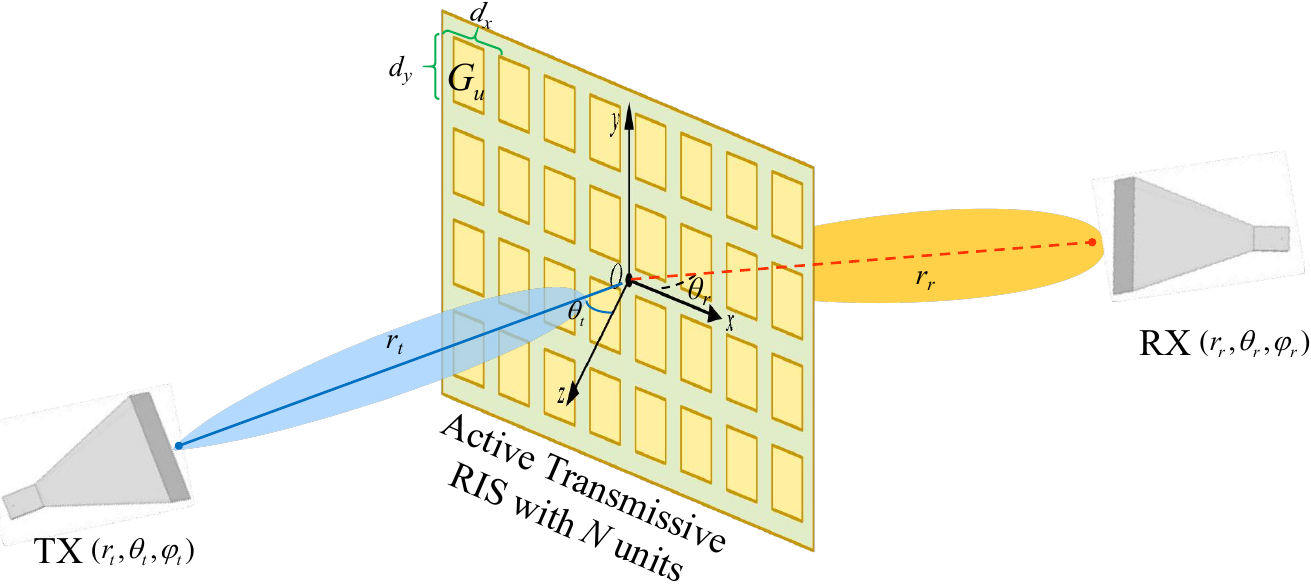}
	\caption{The geometric illustration of a transmissive RIS-aided wireless communication system.}
	\label{fig:system caption}
\end{figure}

As shown in Fig.~\ref{fig:system caption}, the rectangular transmissive RIS, composed of $N = N_x \times N_y$ units, is placed on the $xOy$ plane (i.e., $z = 0$) in a 3D Cartesian coordinate system. The geometric center of the array is located at the origin of the coordinate system, and two sides of the surface are parallel to the $x$-axis and $y$-axis, respectively. 

Let $\boldsymbol{p}_t = \left( x_t, y_t, z_t\right)$, $\boldsymbol{p}_r = \left( x_r, y_r, z_r\right)$ and $\boldsymbol{p}_{n}= \left( x_{n}, y_{n}, z_{n}\right)$ denote the locations of the TX, RX and $n$-th unit in the $n_x$-th row and $n_y$-th column, respectively. Additionally, we use the symbols $r_t$, $\theta_t$, and $\varphi_t$ to denote the equivalent distance, zenith angle, and azimuth angle of the TX relative to the coordinate origin. Similarly, we use the symbols $r_r$, $\theta_r$, and $\varphi_r$ to denote the equivalent distance, zenith angle, and azimuth angle of the RX relative to the coordinate origin. The length and width of the rectangular unit are denoted by $d_x$ and $d_y$, respectively. Therefore, the coordinates of the TX, RX, and $n$-th unit can also be expressed as:
\begin{equation}
    \label{eq:coordinates}
        \begin{array}{l}
            \boldsymbol{p}_t = \left( r_t \sin{\theta_t} \cos{\varphi_t}, r_t \sin{\theta_t} \sin{\varphi_t}, r_t \cos{\theta_t} \right),\\
            \boldsymbol{p}_r = \left( r_r \sin{\theta_r} \cos{\varphi_r}, r_r \sin{\theta_r} \sin{\varphi_r}, r_r \cos{\theta_r} \right),\\
            \boldsymbol{p}_{n}= \left( \delta_{n}^{y} d_x, \delta_{n}^{x} d_y, 0\right),
        \end{array}
\end{equation}
where $\delta_{n}^{y} = n_y - \left( N_y+1\right)/2$ with $n_y \in \left\{1,2,\cdots,N_y \right\}$, and $\delta_{n}^{x} = \left( N_x+1\right)/2 - n_x$ with $n_x \in \left\{ 1,2,\cdots,N_x\right\}$.

Let $\boldsymbol{f} = \left[f_{1}, f_{2}, \cdots, f_{N} \right] \in \mathbb{C}^{1 \times N}$ denote the channel coefficient vector between the TX and the $n$-th unit, where $f_{n} = \alpha_{n} e^{-j\xi_{n}}$ with $n \in \left\{1, 2, \cdots, N\right\}$. Similarly, we denote $\boldsymbol{g} = \left[g_{1}, g_{2}, \cdots, g_{N} \right]^{\text{T}} \in \mathbb{C}^{N \times 1}$ as the channel coefficient vector between the RX and the $n$-th unit, where $g_{n} = \beta_{n} e^{-j\zeta_{n}}$ with $n \in \left\{1, 2, \cdots, N\right\}$. 

Denote the transmit signal and its power by $x$ and $P_t$. The received signal is expressed as
\begin{equation}
    \begin{aligned}
        \label{eq:RX signal}
            y &= \boldsymbol{f} \boldsymbol{\Gamma} \boldsymbol{g} \sqrt{P_t} x + z \\
            &= \left( \sum_{n=1}^{N} h_n g_n \Gamma_n \right) \sqrt{P_t} x + z,
    \end{aligned}
\end{equation}
where $z \sim \mathcal{N}_{\mathbb{C}}\left(0,\sigma_{z}^{2}\right)$ is the additive white Gaussian noise (AWGN) whose variance is $\sigma_{z}^{2}$, and $\boldsymbol{\Gamma} \in \mathbb{C}^{N \times N}$ is the transmission coefficient matrix of units: 
\begin{equation}
    \label{eq:Gamma}
        \boldsymbol{\Gamma} = \text{diag} \left(\mu_{1} e^{j \phi_{1}}, \mu_{2} e^{j \phi_{2}}, \cdots, \mu_{N} e^{j \phi_{N}}\right),
\end{equation}
where $\mu_{n} \in \left[0, 1 \right]$ is the amplitude attenuation caused by the $n$-th unit, and $\phi_{n} \in \left[0, 2\pi \right]$ is the additional phase shift with $i \in \left\{1, 2, \cdots, N\right\}$. 

In free space, the channel coefficient $f_{n}$ in \eqref{eq:RX signal} can be expressed as
\begin{equation}
    \label{eq:hn}
        f_n = \sqrt{\frac{G_t\left(\boldsymbol{\hat{r}}_{n}^{t} \right) A_t\left(\boldsymbol{\hat{r}}_{n}^{t} \right)}{4\pi}}\frac{e^{-j\frac{2\pi}{\lambda}r_n^t}}{r_n^t},
\end{equation}
where $\lambda$ is the wavelength, $G_t\left(\boldsymbol{\hat{r}}_{n}^{t} \right)$ represents the transmit antenna gain, which is related to the direction from the TX to the $n$-th unit; $\boldsymbol{\hat{r}}_{n}^{t}$ is a normal unit vector, $A_t\left(\boldsymbol{\hat{r}}_{n}^{t} \right)$ is the effective area of the $n$-th unit in the TX direction and $r_n^t$ denotes the distance between the TX and the $n$-th unit. 

Similarly, we can obtain the expression of the channel coefficient between the RX and the $n$-th unit by changing the subscript in \eqref{eq:hn} from $t$ to $r$, namely
\begin{equation}
    \label{eq:gn}
        g_n = \sqrt{\frac{G_r\left(\boldsymbol{\hat{r}}_{n}^{r} \right) A_r\left(\boldsymbol{\hat{r}}_{n}^{r} \right)}{4\pi}}\frac{e^{-j\frac{2\pi}{\lambda}r_n^r}}{r_n^r}.
\end{equation}

For the active transmission-type RIS, the signal incident on the unit will first pass through a power amplifier. We use the symbol $G_u\left( u_n\right)$ to denote the gain of each unit where $u_n$ is the voltage (or current equivalently) of the control signal that manipulates the power amplifier. 
Therefore, by substituting \eqref{eq:Gamma} $\sim$ \eqref{eq:gn} into \eqref{eq:RX signal}, we can obtain the detailed model of the received signal: 
\begin{equation}
    \label{eq:detailed RX signal}
        y = \frac{\sqrt{P_t}}{4\pi} \left(
        \sum_{n=1}^{N} \frac{\sqrt{G_t G_r G_u A_t A_r}}{r_n^t r_n^r} \mu_n e^{j \left( \phi_{n} - \Phi_n \right)}
        \right) x + z,
\end{equation}
where $\Phi_n = 2\pi \left( r_n^t + r_n^r \right) / \lambda$ is the phase difference caused by the propagation delay. 
Note that we temporarily drop the arguments of $G_t$, $G_r$, $G_e$, $A_t$ and $A_r$ to make the above expression more concise.

\subsection{A Dual RCS-based Path Loss Model}
\label{subsec:path loss model}
According to \eqref{eq:detailed RX signal}, the received signal power in a transmissive RIS-aided wireless communication system without the consideration of noise can be presented as
\begin{equation}
    \label{eq:RX power}
        P_r = \frac{P_t}{16 \pi^2} \left|
        \sum_{n=1}^{N} \frac{\sqrt{G_t G_r G_u A_t A_r}}{r_n^t r_n^r} \mu_n e^{j \left( \phi_{n} - \frac{2\pi}{\lambda} \left( r_n^t + r_n^r \right) \right)}
        \right|^2.
\end{equation}

In \eqref{eq:RX power}, the amplitude attenuation of the $n$-th unit $\mu_n$ is directly related to the radiation pattern of the unit, which is difficult to measure in practice. Generally, we model it as a trigonometric function. However, it is not applicable to the case of active RIS. To solve this problem, we propose a general path loss model based on the concept of radar cross section (RCS) that is applicable to both active and passive RIS, as well as both reflective and transmissive RIS. 

RCS is a common concept used to describe the scattering efficiency of the target object in the field of antenna design, which is defined as the ratio of the power of the scattered electric field over that of the incident electric field (or the magnetic field equivalently), namely
\begin{equation}
    \label{eq:RCS concept}
        \sigma = \lim \limits_{r \to \infty}4 \pi r^2 \frac{\left| \textbf{E}_s\right|^2}{\left| \textbf{E}_i\right|^2},
\end{equation}
where $r$ is the distance from the receiver radar to the target object, $\textbf{E}_i$ and $\textbf{E}_s$ represent the intensity of the incident electric field and scattered electric field, respectively. It is worth noting that when the transmit and receive radars are in different positions, it is called bistatic RCS.

To make the expression \eqref{eq:RX power} more intuitive, we utilize RCS to depict the joint impact of the amplitude attenuation and the effective area of the transmissive RIS unit, namely
\begin{equation}
    \label{eq:RCS model}
        \sigma_n \left( \boldsymbol{\hat{r}}_{n}^{t}, \boldsymbol{\hat{r}}_{n}^{r}, u_n\right) = \mu_n \sqrt{G_u\left(u_n \right) A_t\left(\boldsymbol{\hat{r}}_{n}^{t} \right) A_r\left(\boldsymbol{\hat{r}}_{n}^{r} \right)}.
\end{equation}
Therefore, the expression \eqref{eq:RX power} can be rewritten as
\begin{equation}
    \label{eq:RCS-based RX power}
        P_r = \frac{P_t}{16 \pi^2} \left|
        \sum_{n=1}^{N} \frac{\sqrt{G_t G_r}}{r_n^t r_n^r} \sigma_n e^{j \left( \phi_{n} - \frac{2\pi}{\lambda} \left( r_n^t + r_n^r \right) \right)}
        \right|^2.
\end{equation}
The corresponding path loss is 
\begin{equation}
    \label{eq:RCS-based path loss}
        PL = \frac{16 \pi^2}{\left|
        \sum_{n=1}^{N} \frac{\sqrt{G_t G_r}}{r_n^t r_n^r} \sigma_n e^{j \left( \phi_{n} - \frac{2\pi}{\lambda} \left( r_n^t + r_n^r \right) \right)}
        \right|^2}.
\end{equation}
For the given positions of TX and RX, \eqref{eq:RCS-based RX power} is maximized as 
\begin{equation}
    \label{eq:max Pr}
        P_{r\max} = \frac{P_t}{16 \pi^2}\left| \sum_{n=1}^{N} \frac{\sqrt{G_t G_r}}{r_n^t r_n^r} \sigma_n \right|^2,
\end{equation}
when 
\begin{equation}
    \label{eq:optimal phi}
        \phi_{n} = \bmod{\left(C + \frac{2\pi}{\lambda} \left( r_n^t + r_n^r \right), 2\pi\right)},
\end{equation}
where $C$ is an arbitrary constant.

Under the assumption of continuous phase shifting capability of RIS, the minimum path loss corresponding to \eqref{eq:max Pr} is 
\begin{equation}
    \label{eq:min path loss}
        PL_{\min} = \frac{16 \pi^2}{\left| \sum_{n=1}^{N} \frac{\sqrt{G_t G_r}}{r_n^t r_n^r} \sigma_n \right|^2}.
\end{equation}

Nevertheless, in practical implementation, obtaining accurate and continuous phase shift $\phi_{n}$ for each unit is difficult due to the actual deployment cost. Instead, the RIS with discrete phase shifts is more commonly designed. For an $m$-bit quantized RIS, the available phase shifts $\phi_{n}$ are
\begin{equation}
    \label{eq:discrete phi}
        \phi_{n} = \left\{0, \frac{1}{2^{m-1}}\pi, \cdots, \frac{2^m-1}{2^{m-1}}\pi \right\} + \psi,
\end{equation}
where $\psi \in \left[0, 2\pi - \frac{2^m-1}{2^{m-1}}\pi \right)$ is an arbitrary angle constant. Then, to maximize \eqref{eq:RCS-based RX power}, beamforming algorithms are needed to solve the problem as follows
\begin{equation}
    \begin{aligned} 
        \label{eq:optimization problem}
            \max_{\phi_{n}} \quad &P_r = \frac{P_t}{16 \pi^2} \left|
        \sum_{n=1}^{N} \frac{\sqrt{G_t G_r}}{r_n^t r_n^r} \sigma_n e^{j \left( \phi_{n} - \Phi_n \right)}
        \right|^2, \\
            \text{s.t.} \quad &\phi_{n} = \left\{0, \frac{1}{2^{m-1}}\pi, \cdots, \frac{2^m-1}{2^{m-1}}\pi \right\} + \psi.
    \end{aligned}
\end{equation}

In this paper, a 2-bit active transmissive RIS is implemented. It is obvious that the actual path loss of our RIS-aided system can not reach the ideal minimum in \eqref{eq:min path loss}. The reasons include imperfect channel state information (CSI) and discrete phase shifts. 

\section{Design of Transmissive RIS Element }

This section presents the detailed design of the phase shift and power amplifier circuit, transmissive RIS units, a 4 $\times$ 4 transmissive RIS array, and a comprehensive performance evaluation of each component. Initially, an evaluation board was developed, featuring a multitude of integrated phase shift and power amplifier devices on a printed circuit board (PCB). For each unit, the receiving and radiating patches are fabricated on the upper and lower surfaces of the PCB, respectively. These two patches are interconnected with a phase shift and power amplifier circuit, which is fold-mounted in the space between patches.

\subsection{Power Amplifying and Phase Shifting Evaluation Board}
\begin{figure}[!htbp]
    \centering 
    \subfigure[]{
    \includegraphics[width=0.4\textwidth]{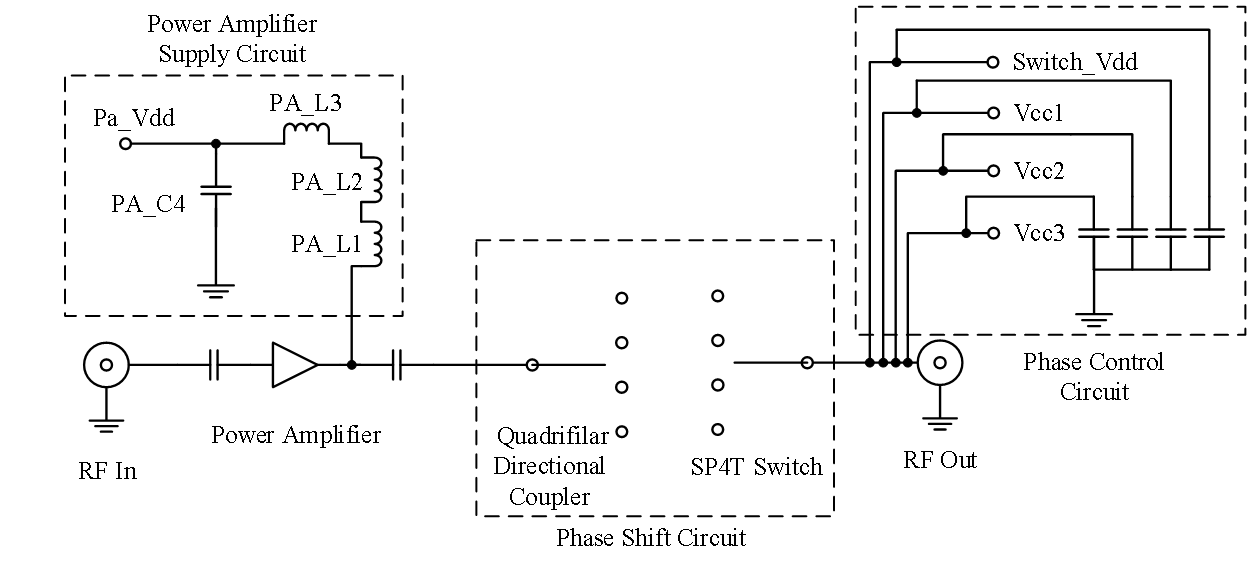}\label{fig:e-board-circuit}}
    \subfigure[]{
    \includegraphics[width=0.33\textwidth]{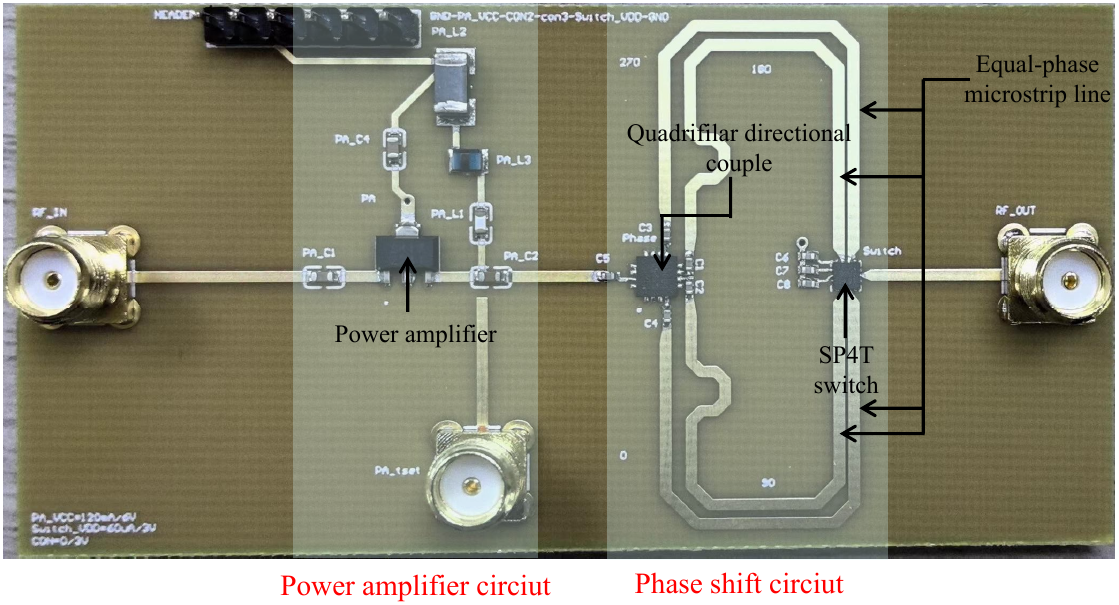}\label{fig:E-board}}
    \caption{Schematic diagram of amplifying phase shift circuit: (a) Schematic of the circuit diagram, (b) Photograph of the fabricated evaluation board.}
    \label{fig:Schematic diagram of e-board}
\end{figure}

To achieve power amplification and phase shifting functionality, the SKY65017-70LF power amplifier from SKYWorks is selected for amplifying the received electromagnetic waves;  the QF2500Q06 quadrifilar directional coupler provides signal outputs with phase shifts of 0$^{\circ}$, 90$^{\circ}$, 180$^{\circ}$, and 270$^{\circ}$. The MXD8641H SP4T switch determines the final output from these different phase signals. Between the coupler and the SP4T switch, four 50$\Omega$ equal-phase microstrip lines establish independent connections between the four signal output ports of the coupler and the four input ports of the SP4T switch. These equal-phase microstrip lines ensure consistent phase delays during signal transmission, thus maintaining the reliability of the 2-bit adjustment. The circuit diagram of the amplifying phase shift circuit is shown in Fig.~\ref{fig:e-board-circuit}.

The evaluation board is fabricated using FR-4 substrate material, with an overall thickness of 0.8 mm and a dielectric constant $\left( \varepsilon_r \right)$ of 4.6. The structure of this board material includes three main layers: the top layer is equipped with a radio frequency circuit layer for signal processing; the middle layer functions as a ground plane, providing electrical stability and reducing electromagnetic interference; and the bottom layer integrates the control signal and power supply layers, ensuring effective power supply and signal control for the entire system.

 An evaluation circuit was designed to verify the cascading reliability among these components, as shown in Fig.~\ref{fig:E-board}. The input signal sequentially passes through a power amplifier, directional coupler, and SP4T switch before output. According to the official technical manual, the SKY65017-70LF power amplifier provides a signal enhancement of 20 dB, effectively compensating for the attenuative impacts induced by the transmission architecture and the various incorporated components within the design. The maximum supply current of the power amplifier is 120 \text{mA}, and the maximum voltage is 5.5 \text{V}. Measured under practical conditions, the maximum gain provided by the power amplifier was 16 dB. The positive supply voltage Vcc is connected to pin 3 of the amplifier through a decoupling network composed of PA\_C4, PA\_L1, PA\_L2, and PA\_L3. The output state of the SP4T switch is determined by three bias control signals (Vcc1, Vcc2, and Vcc3).

 \begin{table}[!htbp]
\renewcommand{\arraystretch}{1.3}
\caption{Control Signal Settings for SP4T switch}
\label{tab:Control_Switched}
\centering
\begin{tabular}{|ccc|c|}
\hline\hline
Vcc1 & Vcc2 & Vcc3 & OUTPUT          \\ \hline
0    & 1    & 1    & 0$^{\circ}$     \\ \hline
0    & 0    & 1    & 90$^{\circ}$    \\ \hline
0    & 0    & 0    & 180$^{\circ}$   \\ \hline
0    & 1    & 0    & 270$^{\circ}$   \\ \hline
\hline
\end{tabular}
\end{table}

\begin{figure}[!htbp]
    \centering 
    \subfigure[]{
    \includegraphics[width=0.4\textwidth]{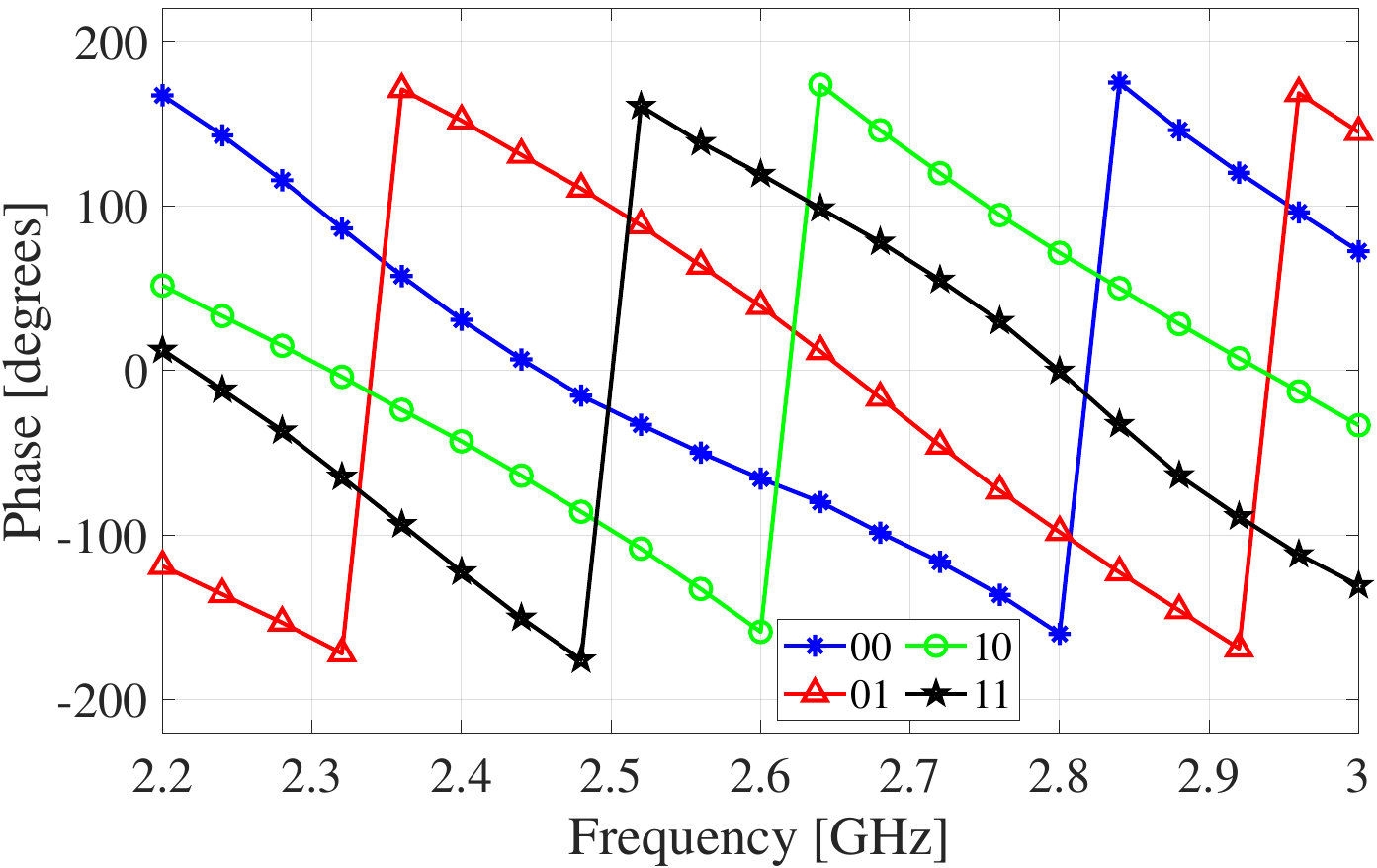}\label{fig:e-phased}}
    \subfigure[]{
    \includegraphics[width=0.23\textwidth]{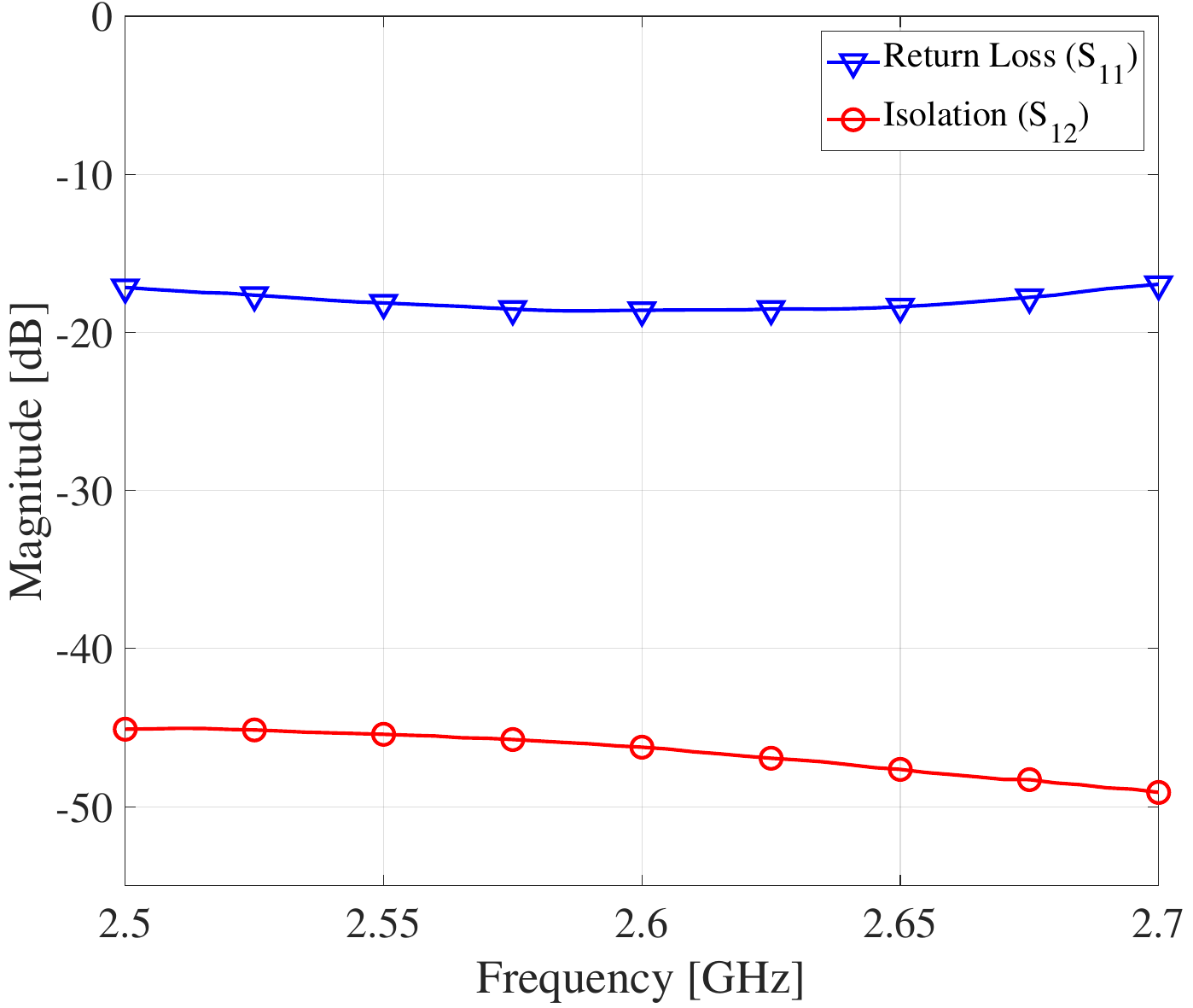}\label{fig:E-iso}}
    \subfigure[]{
    \includegraphics[width=0.23\textwidth]{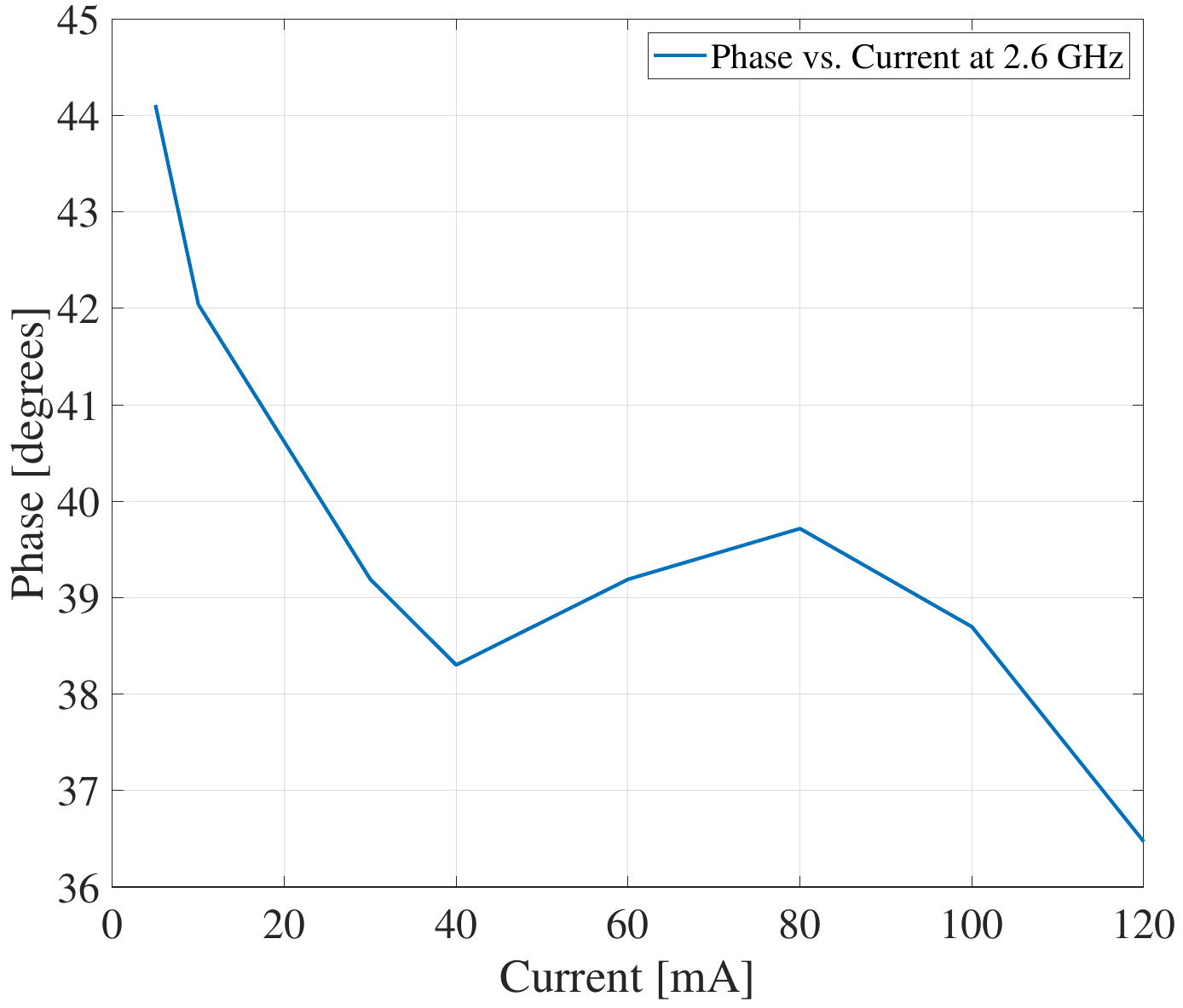}\label{fig:phase_2G6}}
    \caption{Measurement results of evaluation board: phase response and s12 Isolation: (a) The phase response of the evaluation board under four control states, (b) S-Parameter measurement of the evaluation board, (c) Phase at 2.6 GHz under different power amplifier supply currents (different active gains).}
    \label{fig:eboard results}
\end{figure}

 We utilized a phase control method based on SP4T switch control signals. Phase adjustment is achieved by setting the control signal to high (3V, control signals defined as 1) or low (0.3V, control signals defined as 0) while maintaining a constant control current of 0.5µA. The phase adjustment is not directly correlated to the control current yet is governed by the high or low state of the control signal, thus enabling independent control of phase and current. In the experiments, the control signal was initially set to 011, which was defined as the reference state, corresponding to a 0$^{\circ}$ phase of the output signal from the phase shift circuit,  as illustrated in Table~\ref{tab:Control_Switched}. This setting provides a baseline for subsequent phase comparisons. Subsequently, the control signal was adjusted to 001, and it was observed that the phase difference of the output signal was close to 90$^{\circ}$ relative to the reference state, as shown in Fig.~\ref{fig:e-phased}. This result indicates that the phase shift circuit effectively achieved a quarter-cycle phase shift under this control signal. Further, to comprehensively evaluate the performance of the phase shift circuit, the control signal was successively changed to 000 and 010. The measurement results under these settings showed that within the frequency band of interest, the phase difference of the output signal relative to the reference phase was consistently around 180$^{\circ}$ and 270$^{\circ}$. This observation confirms the stability and reliability of the phase shift circuit under different control signals and demonstrates its precision in achieving predetermined phase changes. The measurement results of the evaluation board also show that the phase shift amplification circuit exhibits good return loss and high isolation in Fig.~\ref{fig:E-iso}. At 2.6 GHz, the phase curve at the output of the evaluation board, as shown in Fig.~\ref{fig:phase_2G6}, was obtained by adjusting the supply current of the power amplifier. The amplification and phase shift circuit maintained good phase consistency under different active gain levels, with a phase error of less than 8°.

\subsection{Transmissive RIS Element}
\label{subsec:TRIS Element}
\begin{figure}[!htbp]
    \centering 
    \subfigure[]{
    \includegraphics[width=0.2\textwidth]{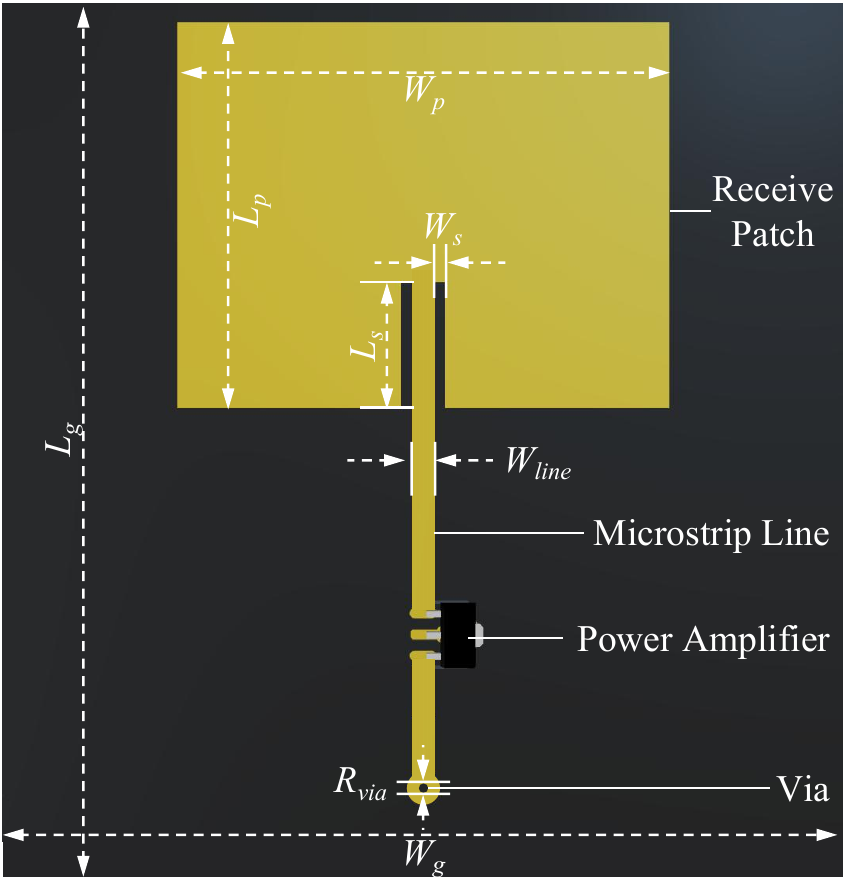}\label{fig:unit top view}}
    \subfigure[]{
    \includegraphics[width=0.21\textwidth]{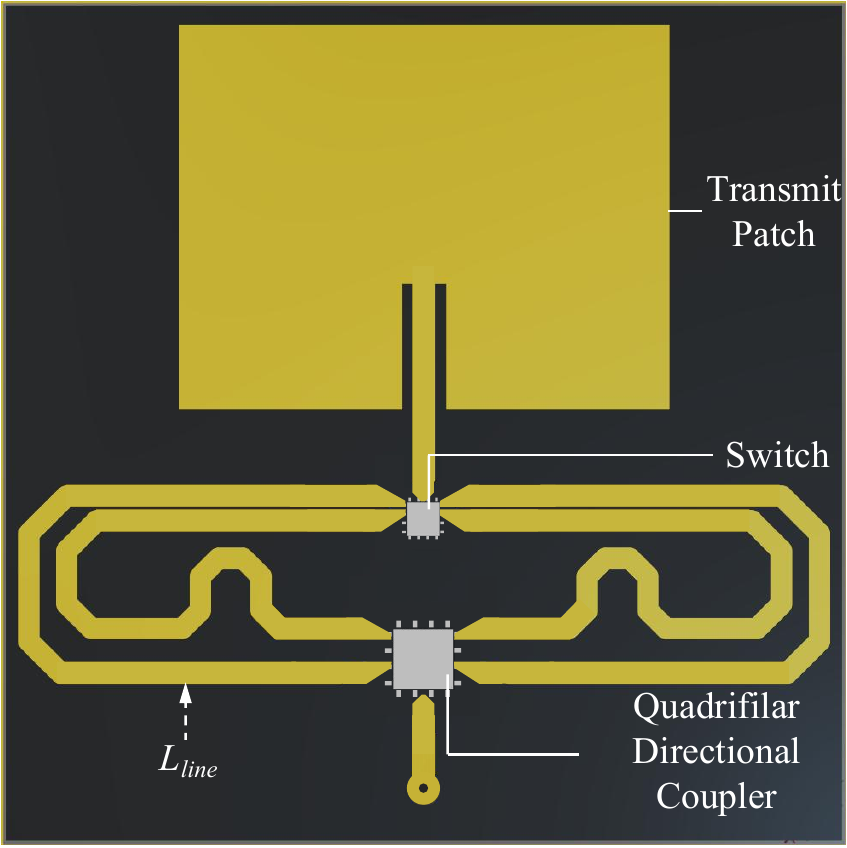}\label{fig:unit bottom view}}
     \subfigure[]{
    \includegraphics[width=0.2\textwidth]{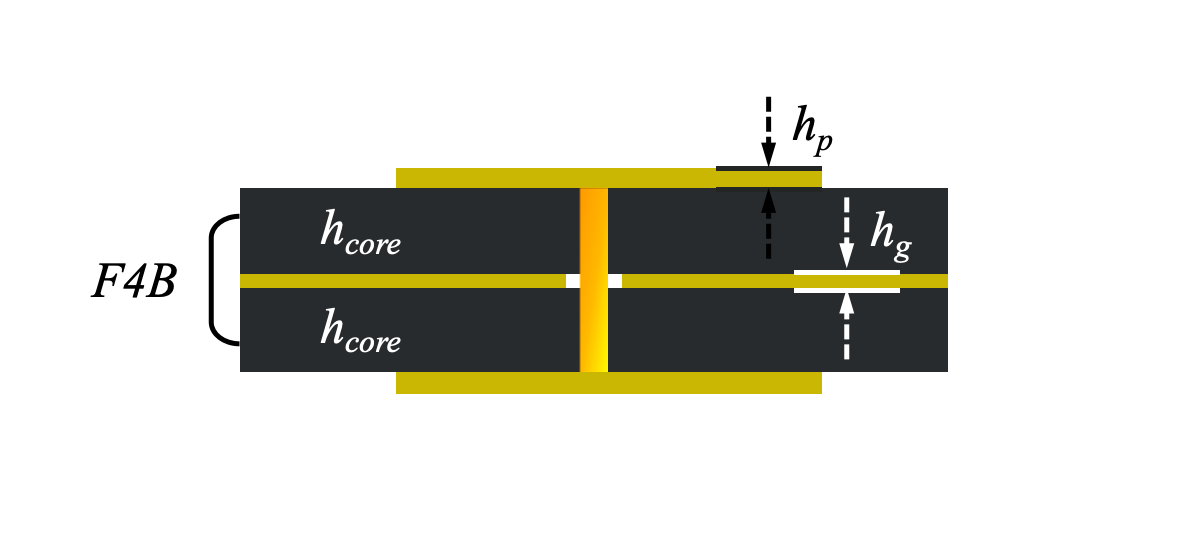}\label{fig:unit lateral view f4b}}
     \subfigure[]{
    \includegraphics[width=0.2\textwidth]{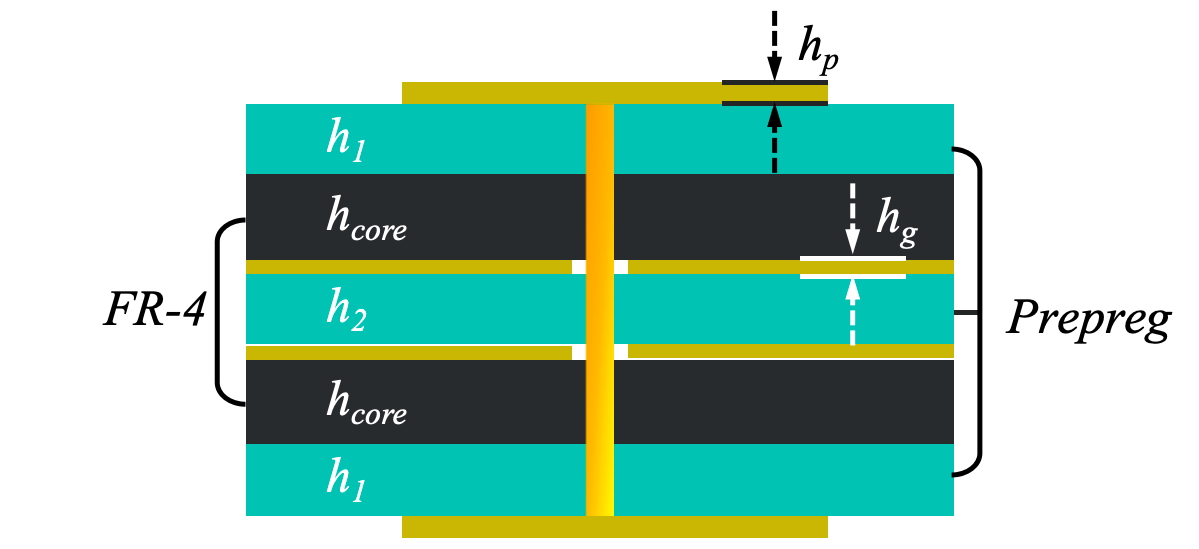}\label{fig:unit lateral view fr4}}
    \caption{Schematic diagram of transmissive RIS unit structure: (a) Top-view. (b) Bottom-view. (c) Lateral-view of the transmissive RIS unit with F4B substrate. (d) Lateral-view of the transmissive RIS unit with FR-4 substrate.}
    \label{fig:Schematic diagram of transmissive RIS unit structure}
\end{figure}

The design of the amplifying transmissive RIS unit structure is illustrated in Fig.~\ref{fig:Schematic diagram of transmissive RIS unit structure}. The dielectric board is constructed using two layers of the F4B substrate, each with 1 mm thick. The size of each element is 60 mm × 60 mm, approximately equal to ${\lambda_{0}/2}$, where $\lambda_{0}$ is the wavelength in free space. This unit comprises two vertically symmetrical common ground patches and a phase shift amplification circuit. The receiving patch on the top layer captures the incident electromagnetic waves. It cascades through a power amplifier to provide signal gain, while the phase shift circuit on the bottom layer supplies adjustable phase signals to excite the patch for radiation, thereby accomplishing the 2-bit phase shift of the electromagnetic wave. The detailed dimensional parameters of this design are presented in Table~\ref{tab:Structural Parameters of RIS Element}. Two patches and components are interconnected by 50 $\Omega$ microstrip lines, with slots cut into the patch to achieve impedance matching. At the terminations of both microstrip lines, a via penetrating the PCB substrate is employed for interconnection.

\begin{table}[!htbp]
\centering
\renewcommand{\arraystretch}{1.3}
\caption{The Structural Parameters of The Transmissive RIS Element.}
\setlength{\tabcolsep}{1mm}{
\label{tab:Structural Parameters of RIS Element}
\centering
\begin{tabular}{|c|c|c||c|c|c|}
\hline\hline
\textbf{Parameter} &\textbf{F4B} &\textbf{FR-4}  &\textbf{Parameter} &\textbf{F4B} &\textbf{FR-4}  \\\hline
L$_\text{g}$       &60 mm        &60 mm          &L$_\text{line}$    &30 mm        &30 mm          \\\hline
W$_\text{g}$       &60 mm        &60 mm          &R$_\text{via}$     &0.4 mm       &0.4 mm         \\\hline
L$_\text{p}$       &35.1 mm      &27.3 mm        &h$_\text{p}$       &35 $\mu$m    &35 $\mu$m      \\\hline
W$_\text{p}$       &42.7 mm      &36.9 mm        &h$_\text{core}$    &1 mm         &0.6 mm         \\\hline
W$_\text{line}$    &2.7 mm       &1.5 mm         &h$_\text{g}$       &35 $\mu$m    &15.2 $\mu$m    \\\hline
L$_\text{s}$       &11 mm        &9.5 mm         &h$_\text{1}$       &——           &0.2104mm       \\\hline
W$_\text{s}$       &1.4 mm       &0.6 mm        &h$_\text{2}$        &——           &0.2028mm       \\\hline
\makecell{$\varepsilon_r$ of\\substrate}         &2.65               &4.6       
&\makecell{$\tan \delta$ of\\substrate}          &0.001              &0.016      \\\hline\hline
\end{tabular}}
\end{table}

\begin{figure}[!htbp]
    \centering
    \includegraphics[width=0.4\textwidth]{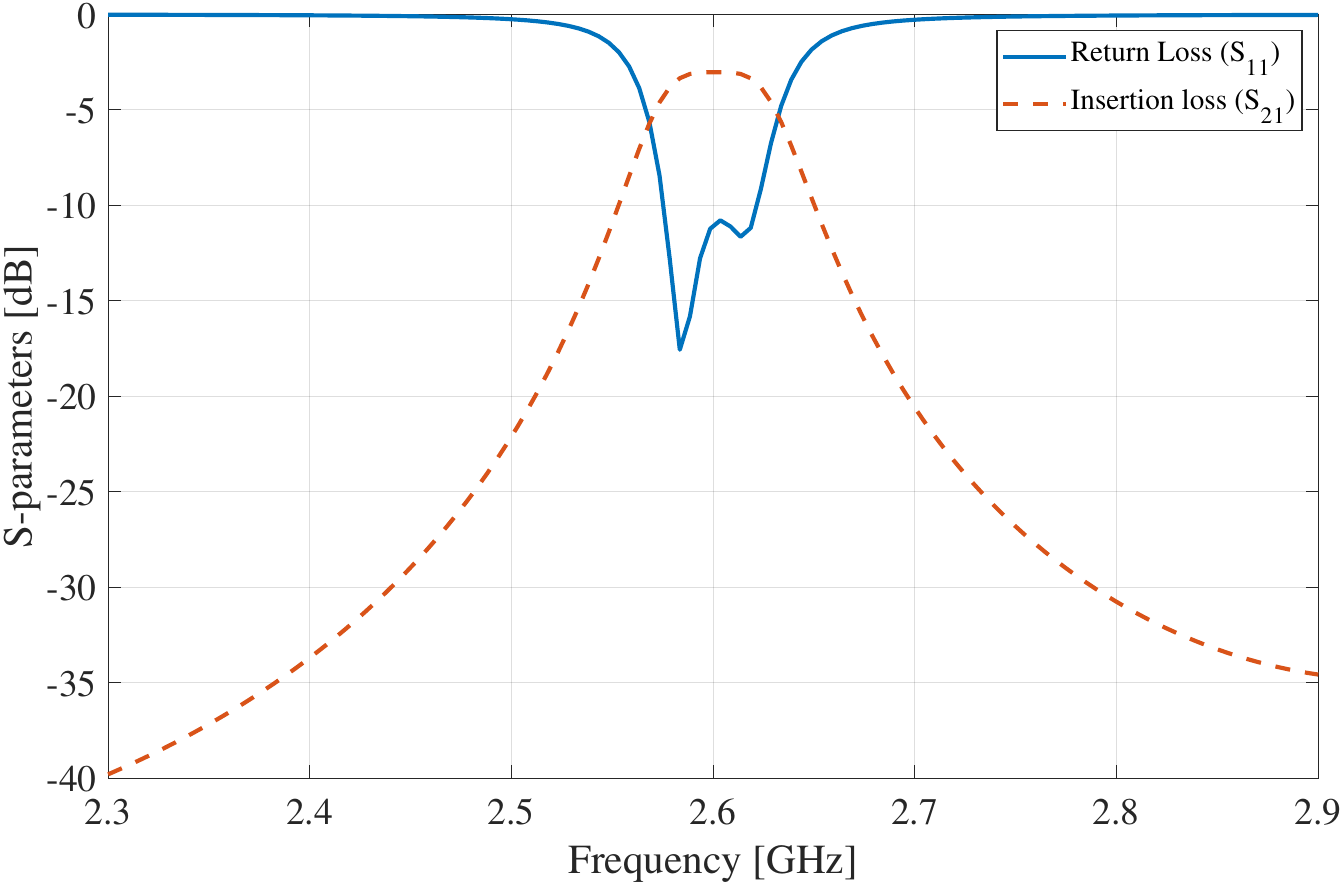}
    \caption{Simulation results of transmissive RIS element with F4B substrate.}
    \label{fig:f4b simulation results}
\end{figure}

The simulations were conducted using Ansys Electronics Desktop 2021 R2. In the simulation of transmissive RIS, master-slave boundary conditions are applied to the four faces immediately surrounding the RIS unit; Floquet ports are utilized on the top and bottom surfaces to enable the simulation of plane wave incidence and reception. The simulation results in Fig.~\ref{fig:f4b simulation results} indicate that within the frequency range of 2.57 GHz to 2.62 GHz, the transmissive RIS consistently maintains a return loss below -10 dB. Simultaneously, the insertion loss remains below -4 dB in this frequency range. These findings suggest a high level of power transmission efficiency and minimal reflection under the specified operational conditions, proving the effectiveness of the design.

To validate the consistency between simulations and actual fabrication, as well as the beamforming capability, active amplification capability, and path loss model reliability of the proposed active transmissive RIS, we fabricated a prototype. Considering the practical aspects of manufacturing costs and process flow, we selected the more cost-effective FR-4 substrate. To accommodate this material change, appropriate simulations were conducted to ensure that the performance of the unit structure with FR-4 matched that initially planned with the F4B substrate. The unit structure was maintained mainly as originally designed, with only some dimensional parameters being optimized and modified to suit the characteristics of the FR-4 material. Details of these structural modifications are presented in Table~\ref{tab:Structural Parameters of RIS Element}. Due to the higher dielectric constant and tangent of the loss angle (tan $\delta$) of the FR-4 substrate compared to the F4B substrate, its electromagnetic losses are more pronounced. This characteristic led to a noticeable degradation in simulation unit performance, particularly in high-frequency applications. While FR-4 substrate offers cost advantages, it compromises electromagnetic performance. For instance, its higher dielectric constant could lead to increased phase delay in signals, and a larger tan $\delta$ (F4B substrate: $\tan \delta$ = 0.001; FR-4 substrate: $\tan \delta$ = 0.016) might result in more energy loss, thus reducing the efficiency and bandwidth of antennas or circuits. Therefore, despite the cost and processing benefits of FR-4 substrate, careful consideration must be given to the trade-offs in electromagnetic performance it entails. We optimize this unit for better performance while reducing costs. 

\begin{figure}[!htbp]
    \centering
    \includegraphics[width=0.4\textwidth]{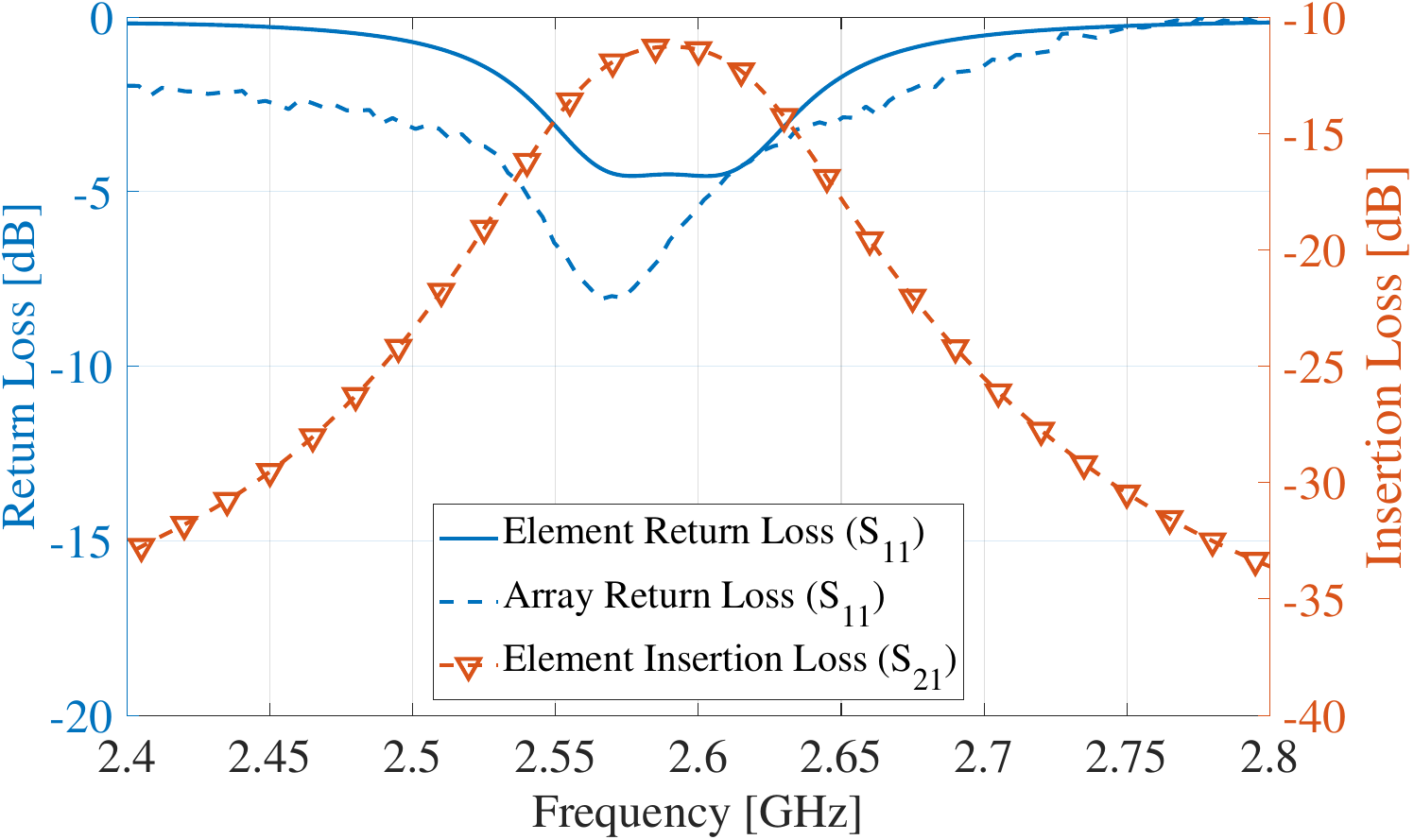}
    \caption{Simulation results of transmissive RIS element and $4 \times 8$ array with FR-4 substrate.}
    \label{fig:fr4 simulation results}
\end{figure}

After optimization, we obtain the simulation results shown in Fig.~\ref{fig:fr4 simulation results}. These results also confirm the significant differences in electromagnetic performance between FR-4 and F4B substrates. Specifically, in terms of return loss, we observed a performance degradation of approximately -10 dB when using the FR-4 substrate compared to the F4B substrate. Regarding insertion loss was approximately -11 dB.

\subsection{Duplex Transmissive RIS Element}

In the practical deployment of RIS systems, particularly transmissive RIS systems, it is crucial not only to amplify and control the downlink signals from the base station to the user but also to ensure efficient transmission of the reverse uplink signals. However, in traditional active RIS designs, the performance of reverse signal transmission is compromised due to the high isolation typically exhibited by power amplification circuits. Building on the active transmissive RIS design in Sec. \ref{subsec:TRIS Element}, we have introduced modifications aimed at enhancing the duplex functionality of RIS system.

\begin{figure}[!htbp]
    \centering 
    \subfigure[]{
    \includegraphics[width=0.3\textwidth]{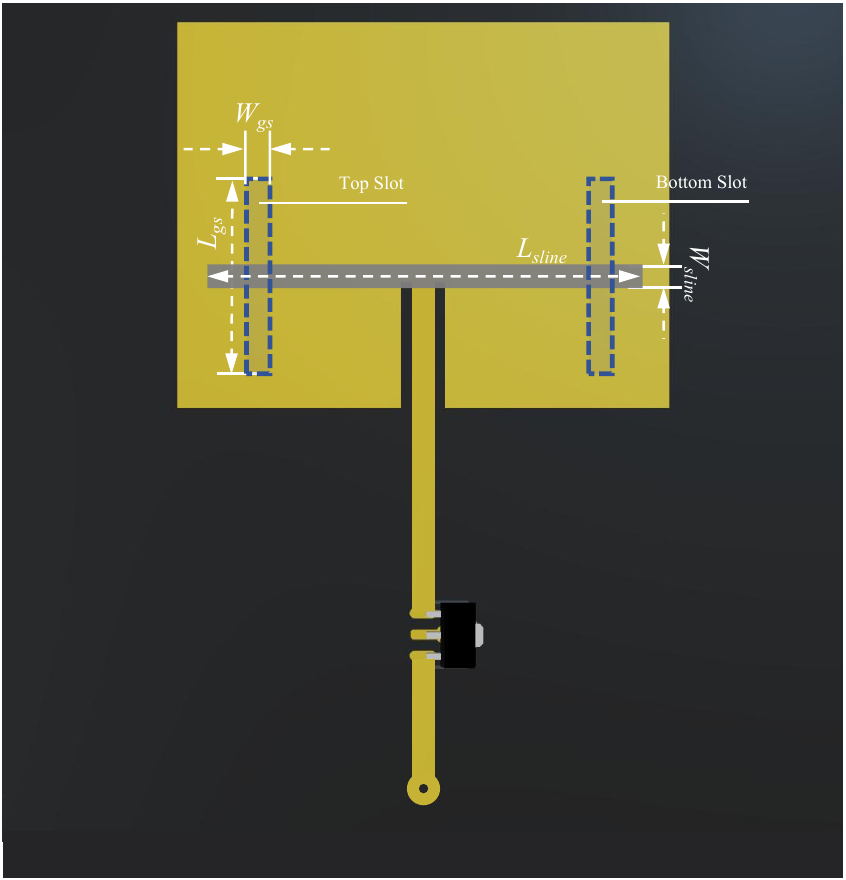}\label{fig:duplex unit top view}}
    \subfigure[]{
    \includegraphics[width=0.3\textwidth]{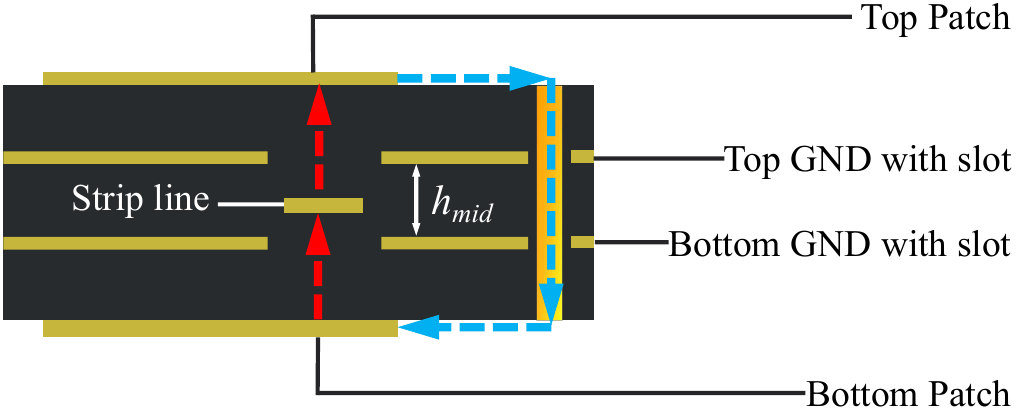}\label{fig:deplex unit lateral view}}
    \caption{Schematic diagram of duplex transmissive RIS unit structure: (a) Top-view. (b) Lateral-view. (L$_\text{gs}$=17.9 mm, W$_\text{gs}$=2 mm, L$_\text{sline}$=26.6 mm, W$_\text{sline}$=0.65 mm, h$_\text{mid}$=1 mm).}
    \label{fig:Schematic diagram of Duplex RIS}
\end{figure}

The overall structure of the duplex transmissive RIS element is highly comparable to that of the non-duplex RIS design utilizing F4B substrate, with several critical modifications. These modifications include the incorporation of slots in both the top and bottom ground planes, the insertion of an F4B dielectric substrate with an overall thickness of 1 mm between the two ground planes, and the integration of a stripline transmission line centrally positioned within the dielectric substrate. These design enhancements are intended to facilitate the realization of an active transmissive RIS based on Frequency Division Duplex (FDD) mode. The structural schematic of the unit is shown in Fig.~\ref{fig:Schematic diagram of Duplex RIS}. In this structure, the transmission path of the forward incident wave sequentially passes through the top patch, the phase-shifting and amplification circuit, and the via connecting the top and bottom layers, ultimately being transmitted through the bottom patch. This pathway is represented by the blue dashed lines in Fig.~\ref{fig:deplex unit lateral view}. In contrast, the reverse incident wave follows a different transmission path, starting at the bottom patch, where it initially receives the reverse wave. The wave then progresses through the slot on the bottom ground plane, the stripline located at the center of the structure, and the slot on the top ground plane, finally being transmitted through the top patch. This process is depicted by the red dashed lines in Fig.~\ref{fig:deplex unit lateral view}.

\begin{figure}[!htbp]
    \centering
      \subfigure[]{
    \includegraphics[width=0.4\textwidth]{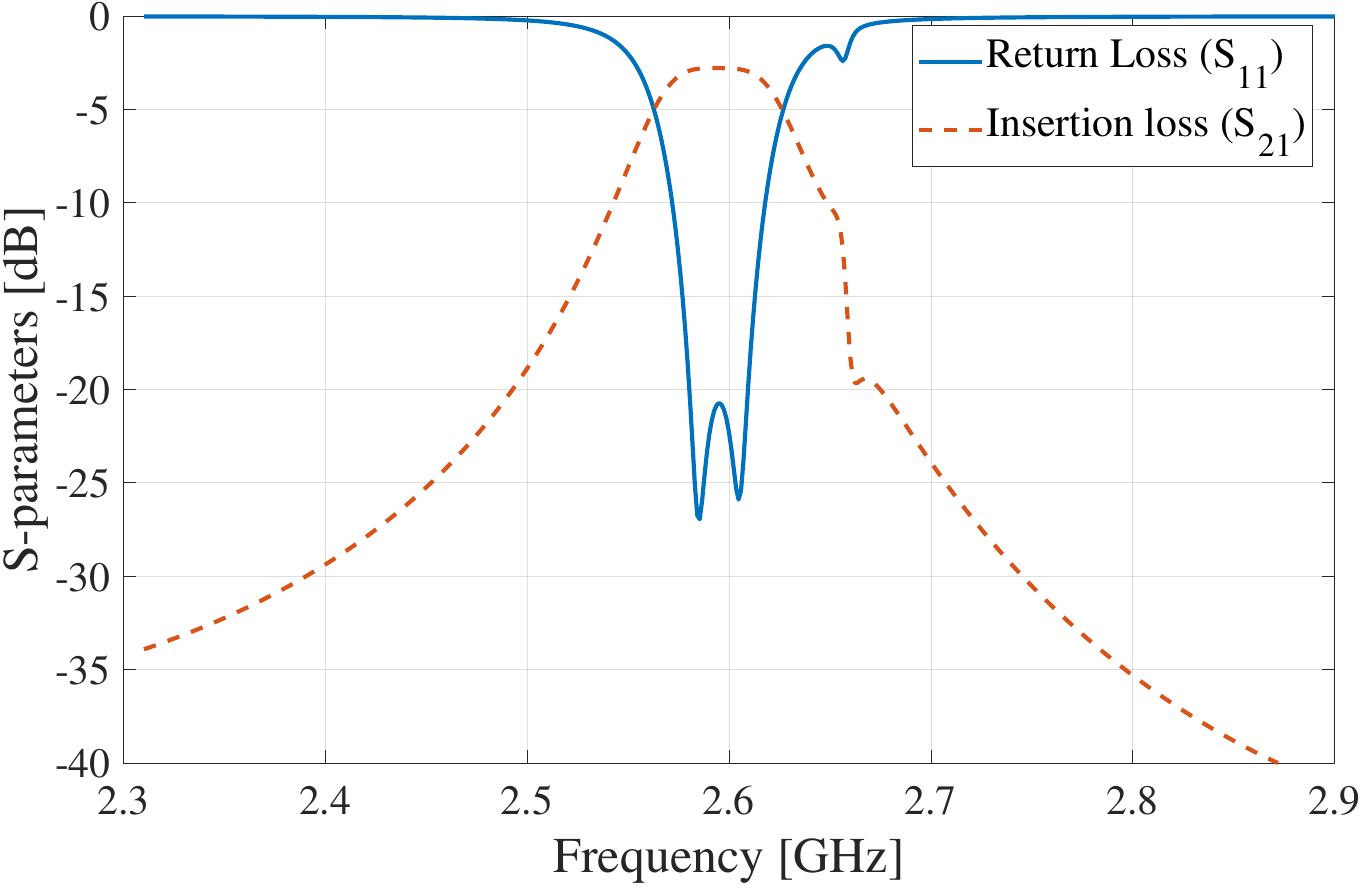}\label{fig:duplex downlink mode}}
    \subfigure[]{
    \includegraphics[width=0.4\textwidth]{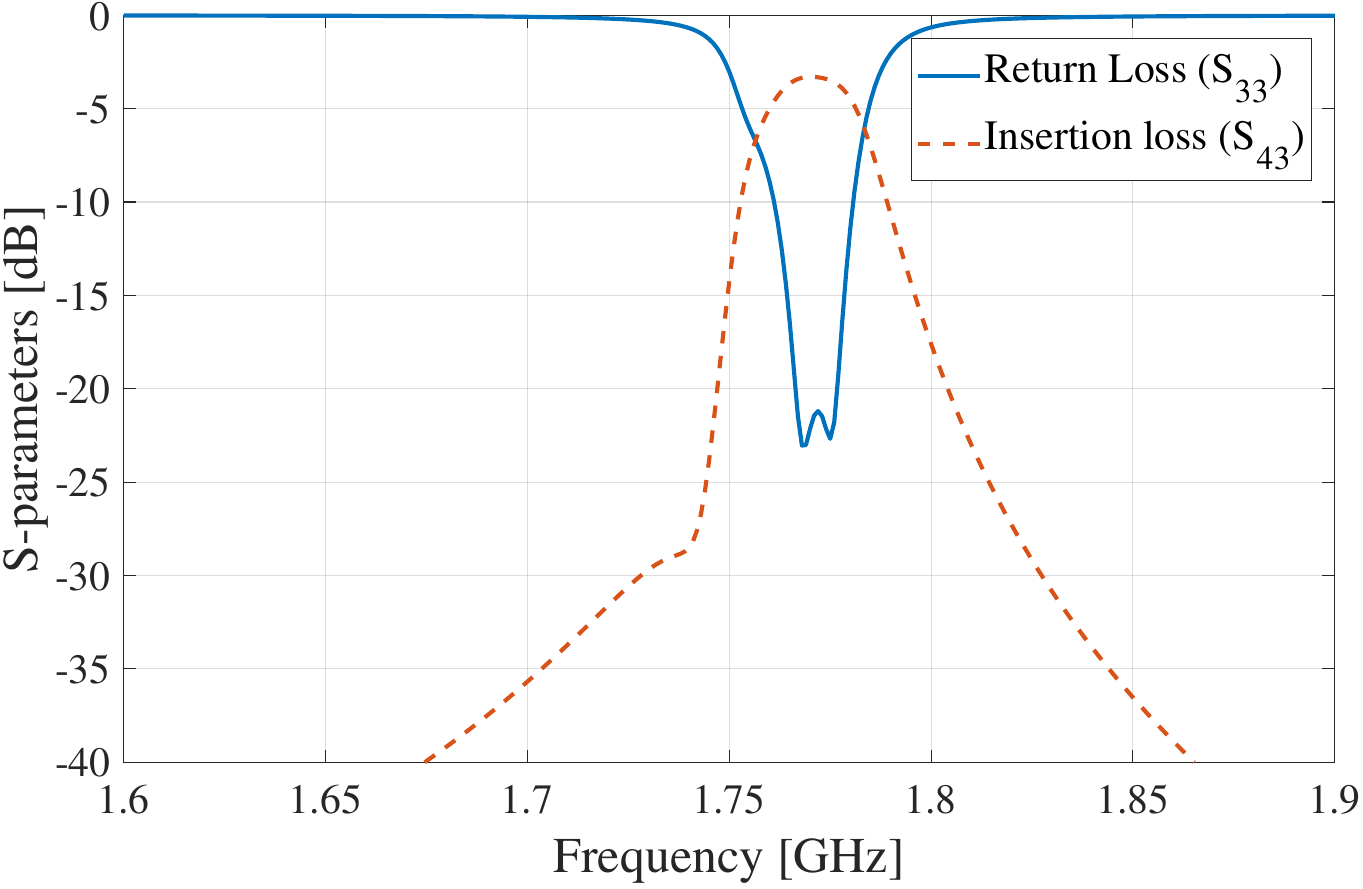}\label{fig:deplex uplink mode}}
    \caption{Simulation results of the duplex transmissive RIS element: (a) Downlink mode, (b) Uplink mode.}
    \label{fig:Duplex simulation results}
\end{figure}
Through these structural adjustments, the RIS unit is capable of efficiently transmitting uplink signals at the 1.76 GHz frequency, which, like the 2.6 GHz band, is also part of the commercial communication spectrum utilized by China Mobile, thereby enabling duplex functionality. Moreover, the polarization of the reverse incident electromagnetic waves is orthogonal to that of the forward incident waves, ensuring high isolation between them and thus facilitating the implementation of full-duplex operation. Based on the simulation results, it is evident that during reverse transmission of the duplex RIS element, the insertion loss remains better than -4 dB. The simulation results of the duplex transmissive RIS Element are shown in Fig.~\ref{fig:Duplex simulation results}.

\section{Measurements and Verification}

\subsection{Design Details of Transmissive RIS Array}
\begin{figure}[!htbp]
    \centering 
    \subfigure[]{
    \includegraphics[width=0.2\textwidth]{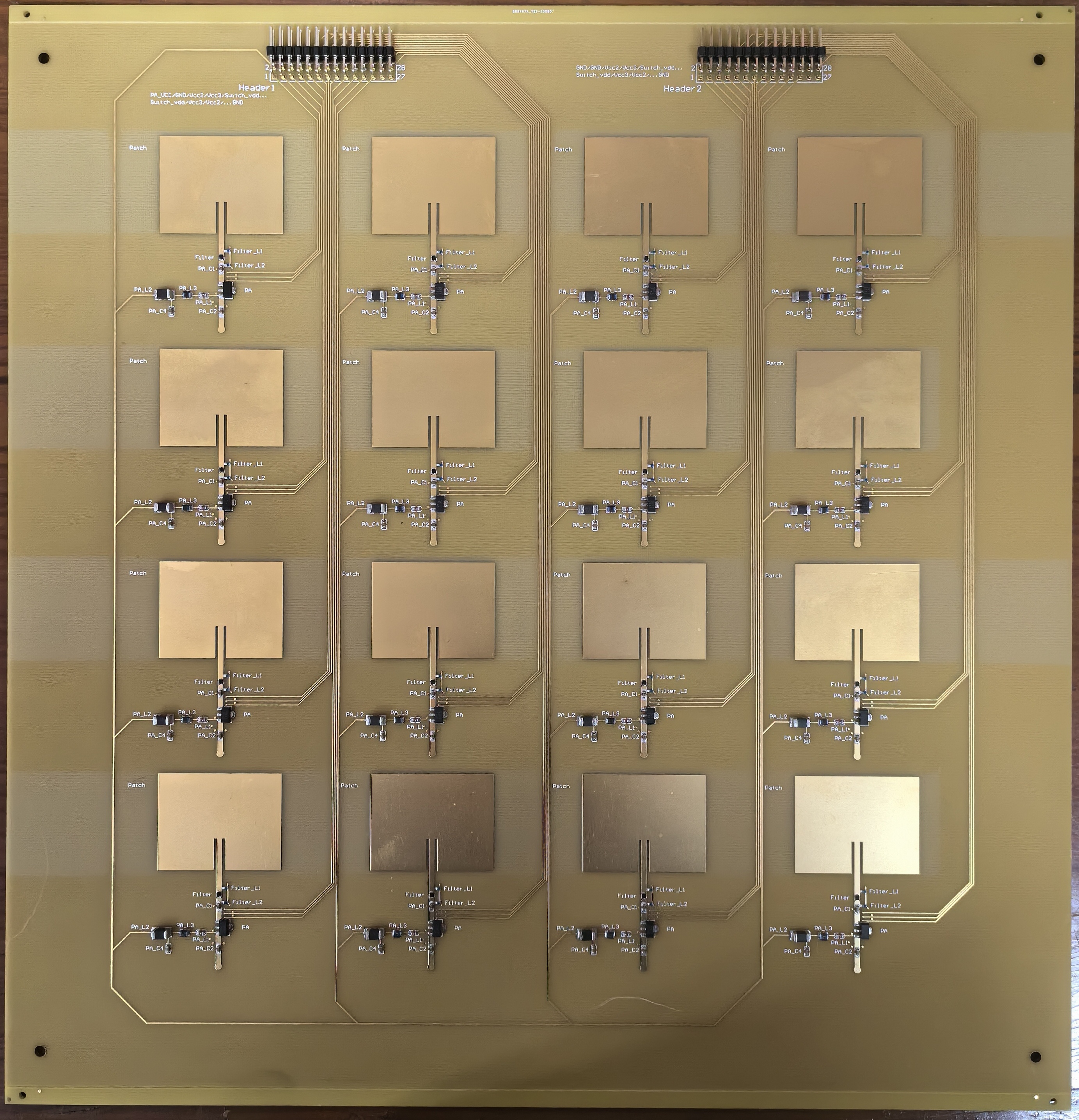}}
    \subfigure[]{
    \includegraphics[width=0.2\textwidth]{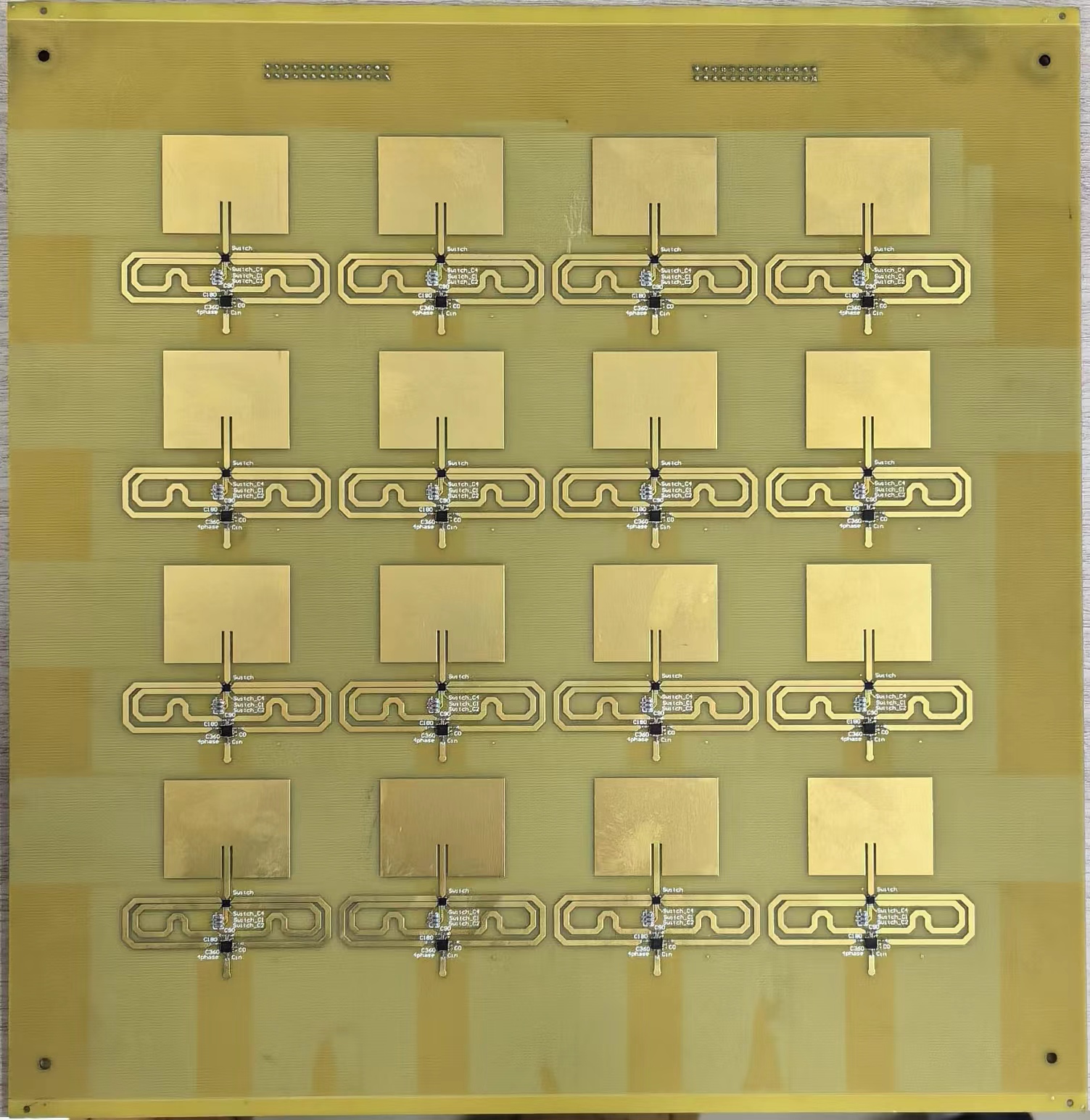}}
    \caption{Fabricated transmissive RIS with FR-4 substrate. (a) Top view. (b) Bottom view.}
    \label{fig:transmissive RIS array}
\end{figure}

The design of the transmissive RIS array is depicted in Fig.~\ref{fig:transmissive RIS array}. The overall structure of the array is a 310 $\times$ 310 \text{mm} FR-4 printed circuit board featuring a laminated board structure consistent with that used in the unit design. This structure comprises two layers of 0.8 \text{mm} thick FR-4 substrates. They are tightly bonded together through a lamination process. The array is divided into upper and lower sections, totaling 16 independent transmissive RIS units. Each unit is equipped with independent control circuits, encompassing phase control and power amplification supply circuits. These circuits are connected to two bullhorn sockets at the top of the array, facilitating communication through ribbon cables with the control board at the rear. This design allows for independent adjustment of gain and phase amplitude for each unit within the entire transmissive RIS array.

\subsection{Beamforming Algorithm for Tansmissive RIS}
\label{subsec:beamforming algorithm}
A fast-beamforming algorithm with low training/feedback overhead previously proposed in \cite{pei2021ris}  has been applied in the current measurement of the transmission array. 
This algorithm, functioning as a blind beamforming technique, obviates the need for additional channel information or user location. Instead, it adjusts the phase configuration solely based on the feedback of the quality of the received signal. Particularly suited for complex and dynamic communication environments, this algorithm can rapidly respond to environment changes and adapt its beamforming coefficients accordingly.

This algorithm aims to optimize the signal reception quality in wireless communication systems by adjusting the transmission coefficients of RIS. The process begins with initializing an \( M \times N \) complex matrix \( T_0 \), representing the initial state of the RIS transmission coefficients. Following this initialization, the algorithm receives initial feedback on the receiver (RX) signal power \( p_0 \), establishing a baseline for subsequent comparisons and adjustments.

The algorithm then proceeds in two main steps: horizontal search and vertical search.

\begin{itemize}
\item Horizontal Search: This step focuses on each column of the RIS matrix. For each column (from the 1st to the Nth), the algorithm attempts to optimize the received signal power by altering the phase of the current column. Specifically, the algorithm increases the phase of the current column by 90$^{\circ}$  and then receives feedback on the signal power under the new configuration \( p_n \). If this new configuration maintains or improves the received power (\( p_{n-1} \geq p_n \)), the change is retained. Otherwise, the change is reverted, maintaining the previous configuration.
    
\item Vertical Search: After completing the horizontal search, the algorithm focuses on each row of the RIS matrix. This process is similar to the horizontal search yet operates on rows instead of columns. Similarly, the algorithm attempts to find a more optimal transmission coefficient configuration by changing the phase of each row, thereby enhancing the received signal power.
\end{itemize}

Through these iterative steps, the algorithm gradually optimizes the transmission coefficient configuration of the entire RIS array. As the algorithm adjusts only one row or column at a time, it incrementally improves the signal-to-noise ratio (SNR) of the system without causing significant fluctuations in the received power. This step-wise iterative approach has a very low complexity. 


\subsection{Phase and Amplitude Response Measurement}
\label{subsec:phase and amplitude measurement}

\begin{figure}[!htbp]
    \centering
    \includegraphics[width=0.35\textwidth]{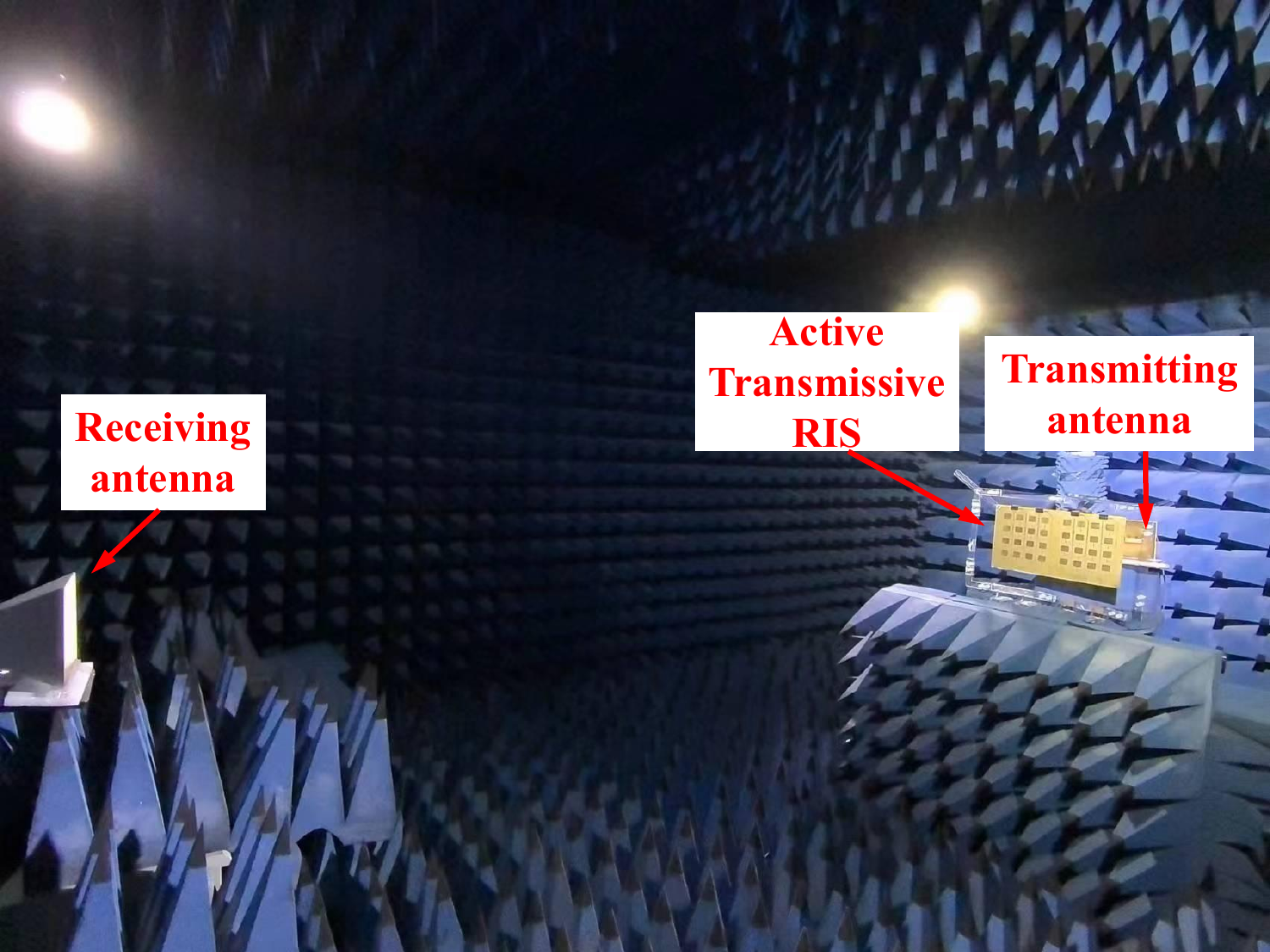}
    \caption{Transmissive RIS array measurement environment.}
    \label{fig:environment}
\end{figure}

In this section, we assess the phase control capabilities and signal amplification efficiency of the proposed transmissive RIS array. The experiments were conducted in a microwave anechoic chamber, the measurement setup consists of two standard gain horn antennas, used respectively as the transmitter and receiver. A $4 \times 8$ transmissive RIS array with 32 units was precisely positioned on the central turntable within the chamber, allowing for multi-angular and directional measurements. The transmitting horn antenna was placed at the exact center on the receiving face of the transmissive RIS array, maintaining a distance of 60 \text{cm} to ensure perpendicular incidence and good beam coverage. The receiving horn antenna was aligned at the same horizontal level as the transmissive RIS array and the receiver and is positioned perpendicular to the radiation face of the RIS array. Furthermore, the distance between the receiver and the transmissive RIS array was 4 \text{m}. A detailed schematic of this measurement environment is provided in Fig.~\ref{fig:environment}. This designed measurement setup enabled us to comprehensively and accurately assess the performance of the transmissive RIS array in a controlled and standardized environment, particularly focusing on the key technical indicators of phase control and signal amplification.

\begin{figure}[!htbp]
    \centering 
    \subfigure[]{
    \includegraphics[width=0.325\textwidth]{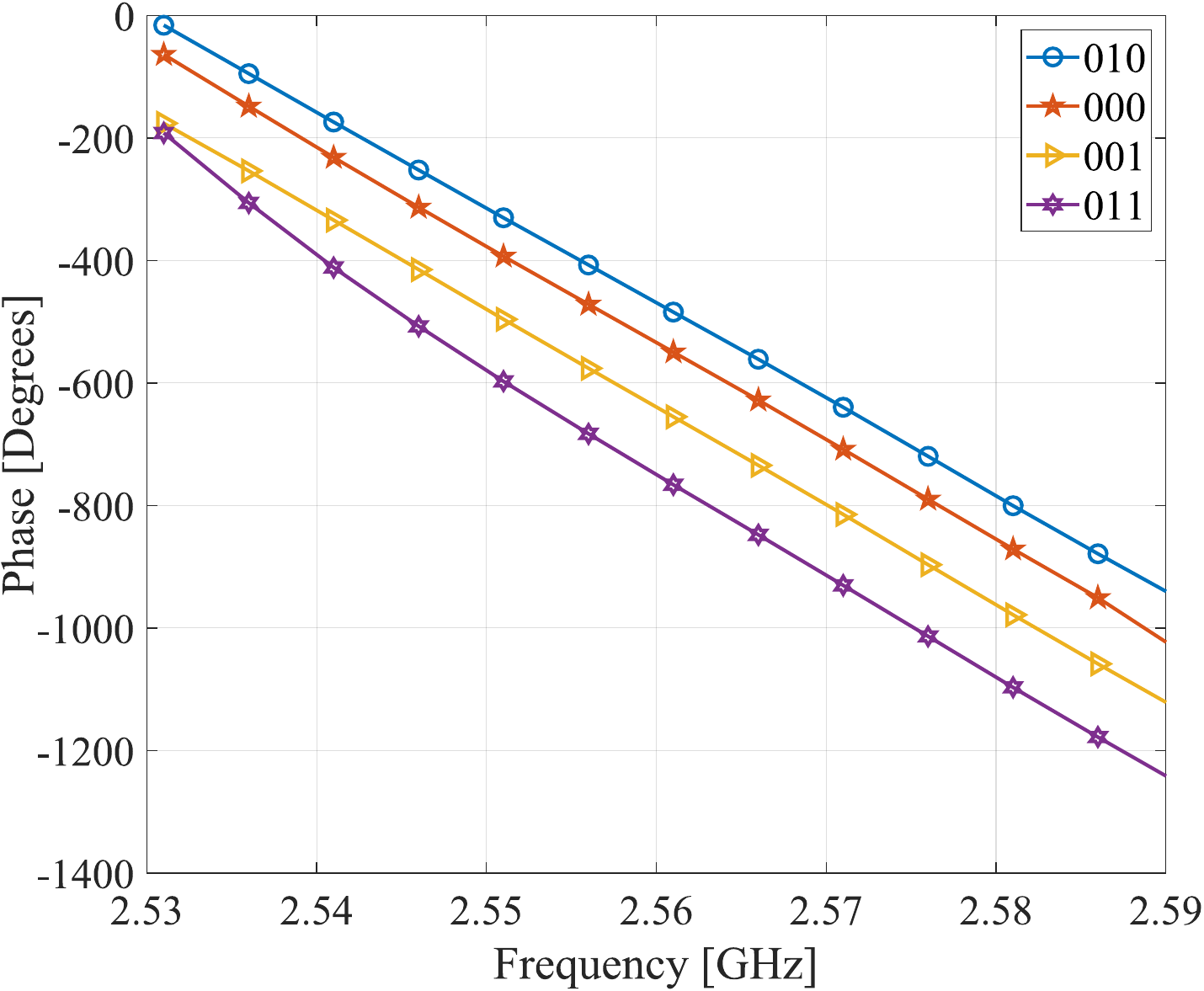}}
    \subfigure[]{
    \includegraphics[width=0.3\textwidth]{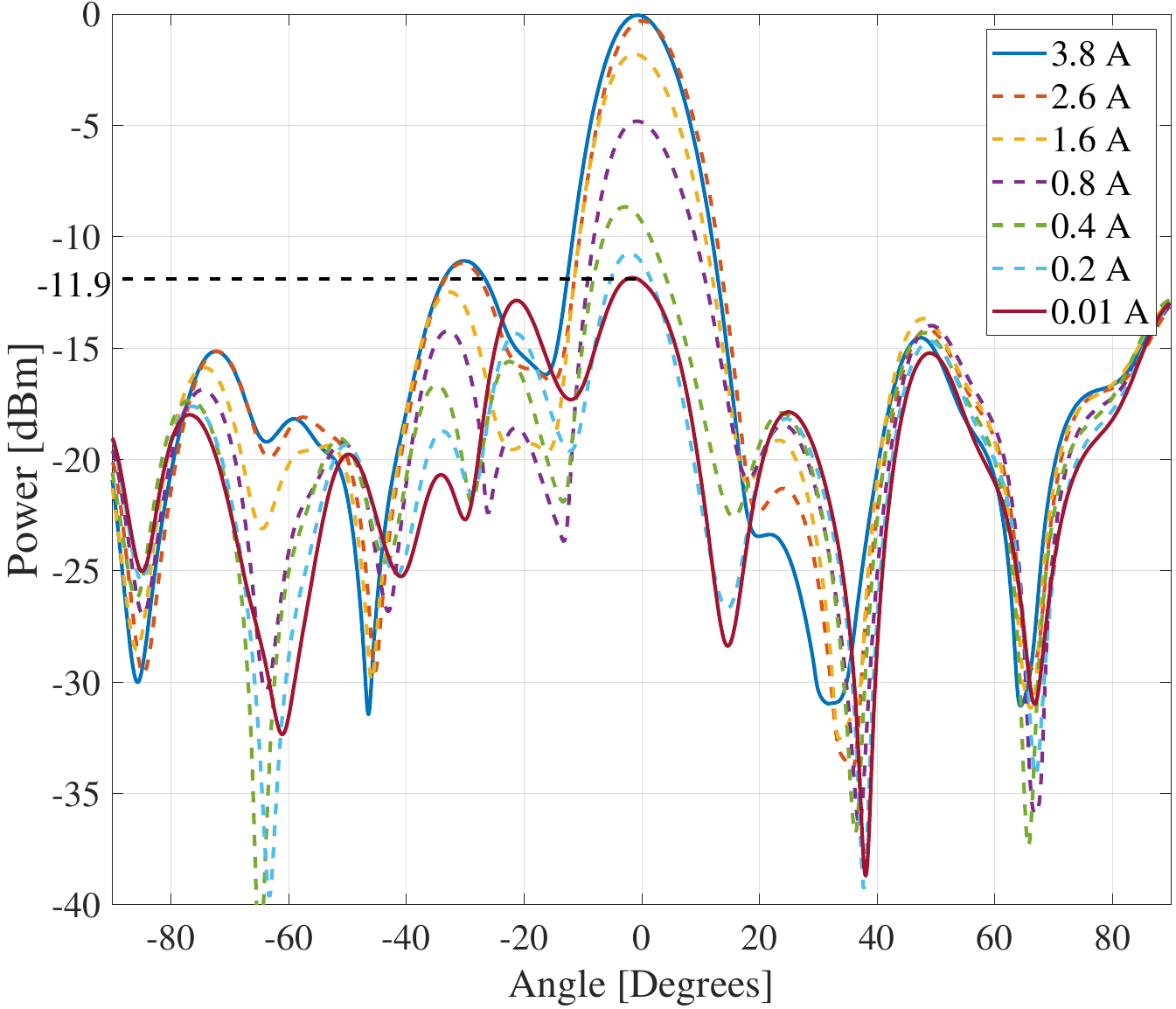}}
    \caption{Transmissive RIS array phase and amplitude response measurement results. (a) Phase response under four control states. (b) Amplitude response under various supply currents.}
    \label{fig:0-PA-13}
\end{figure}

We initially set the transmissive RIS array to operate at full power to ensure an accurate assessment of its capabilities. We sequentially switched the control signals of the selection switches for each unit, achieving a uniform phase change across the entire array. A vector network analyzer was placed outside the microwave anechoic chamber to capture phase information precisely. As demonstrated in  Fig.~\ref{fig:0-PA-13}(a), by sequentially toggling through four distinct states within the 2.55-2.57 GHz frequency range, the four particular states demonstrated phase differences closely approximating 90$^{\circ}$ each. This process validated the precision of the fabricated transmissive RIS in phase control.

To analyze the variations in the radiation pattern of the RIS under different active gain levels, we adjust the reception angle of the receiver to 0° and then execute the beamforming algorithm to steer the beam toward the receiver.
By precisely controlling the input power, we recorded the response characteristics of the received sAignal at each power level. Subsequently, we measured the directional patterns of the RIS array for each power state, as shown in Fig.~\ref{fig:0-PA-13}(b). The measurement outcomes reveal a pivotal aspect of the capabilities of the transmissive RIS array. Specifically, power attainment at the receiver peaked when the supply current for the power amplifier chip was 1.4 \text{A}. This peak represents a significant gain of 11.9 \text{dB} in received power when contrasted with the minimum supply current of 0.01 \text{A}. 
Concurrently, this finding also substantiates the proficiency of the transmissive RIS in amplifying incident signals, a distinctive attribute not inherent in traditional RIS systems. As shown in Fig.~\ref{fig:0-PA-13}(b), when the power amplifier operates with a high current, providing significant active gain, the radiation pattern exhibits good directionality, with the maximum gain of the main lobe decreasing in sync with the active gain. This ensures signal strength and coherence, maintaining good beam directionality. However, when the operating current is low, such as at 0.2A and 0.01A, the array pattern changes significantly because the reduction in active gain weakens beam coherence. Higher active gain helps maintain coherence, effectively focusing energy on the main lobe. Lower gain can cause beam dispersion, changing the main lobe shape and relatively increasing sidelobe levels. Additionally, lower active gain makes the system more susceptible to nonlinear effects, further impacting beam shape and sidelobe levels. 

\subsection{Validations of the Dual RCS-based Path Loss Model}

\begin{figure}[!htbp]
	\centering
	\includegraphics[width=0.33\textwidth]{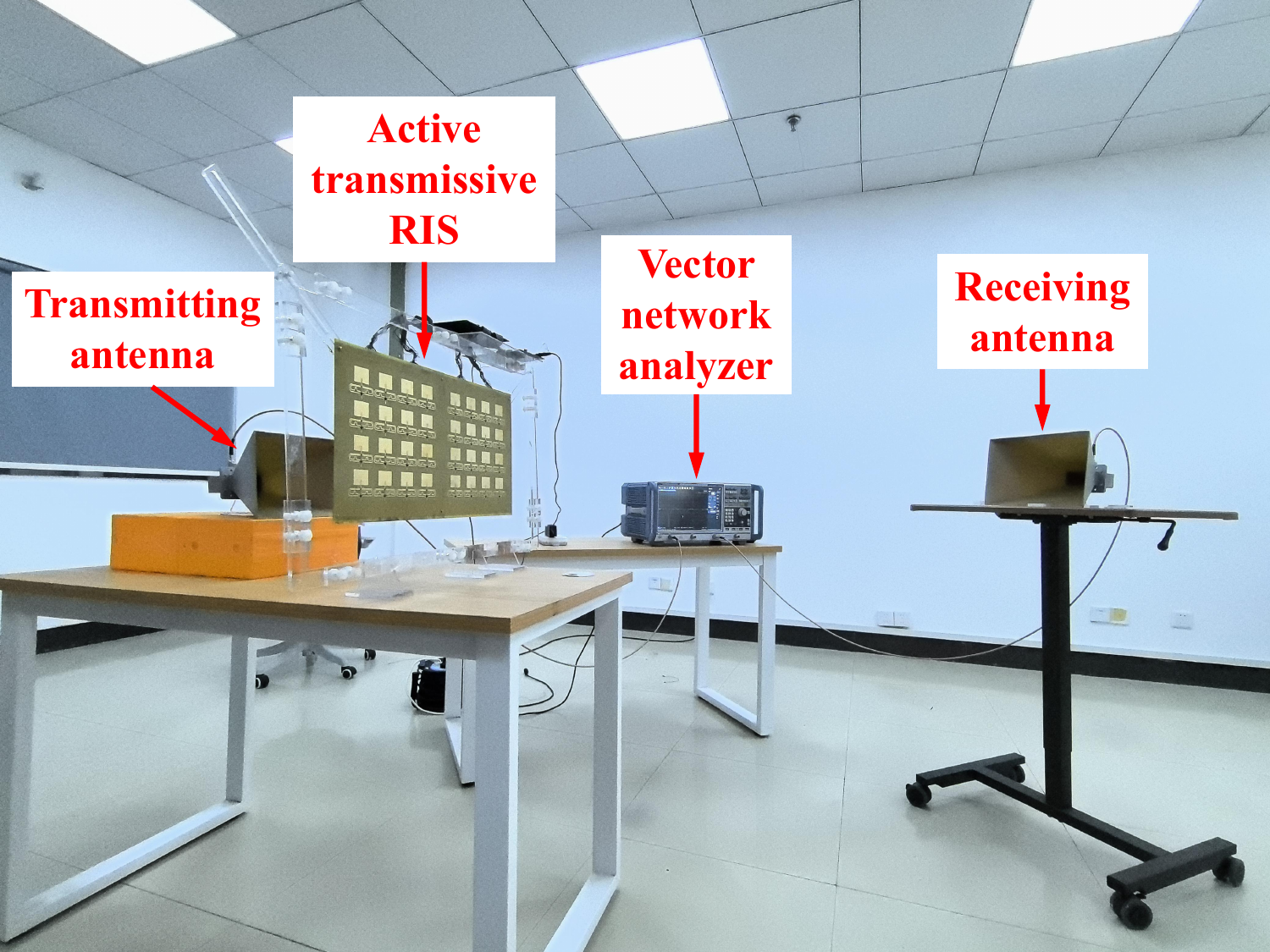}
	\caption{Path loss validation setup for transmissive RIS array in laboratory environment}
	\label{fig:Lab-Environment}
\end{figure}

In section \ref{subsec:path loss model}, we proposed a dual RCS-based path loss model \eqref{eq:RCS-based path loss}. From this model, we observe that the path loss increases approximately quadratically with the product of the TX-RIS distance and the RIS-RX distance. Moreover, the path loss is related to the angles of TX/RX, the size of RIS, the regulation ability of RIS, etc. To validate the effectiveness of the model, we carried out experiments in the laboratory environment for distances and in an anechoic chamber for angles illustrated in Fig. \ref{fig:Lab-Environment} and Fig.  \ref{fig:environment}, respectively. The size of the transmissive RIS was $4 \times 8$, i.e., 32 units and two standard gain horn antennas were adopted to play the role of transmitter and receiver, respectively.

\begin{figure}[!htbp]
	\centering
	\includegraphics[width=0.33\textwidth]{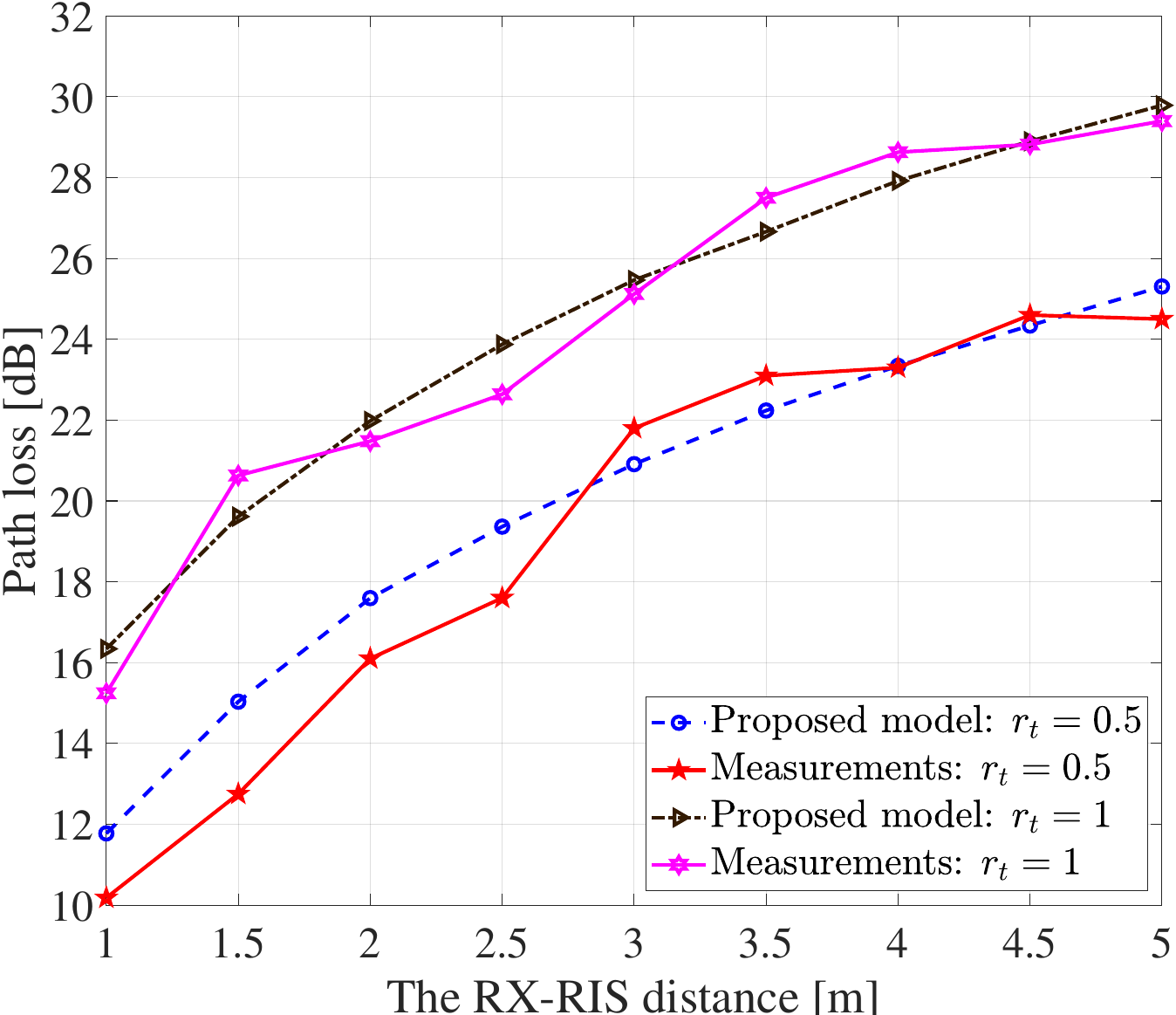}
	\caption{Path loss versus the RX-RIS distance $r_r$ with $\theta_{t} = 0, \theta_{r} = 0, \varphi_t = 0, \varphi_r = 0$.}
	\label{fig:plot_dis}
\end{figure}

To begin with, we study the impact of distances on the path loss. The transmitting horn antenna was placed perpendicular to the array plane facing the geometric center of the RIS array in one side with a distance of 50 \text{cm} or 100 \text{cm}, namely case 1: $r_t = 0.5\; \text{m},\;\theta_{t} = \varphi_t = 0^{\circ}$ and case 2: $r_t = 1\;  \text{m},\;\theta_{t} = \varphi_t = 0^{\circ}$. Similarly, the receiving horn antenna was placed perpendicular to the array plane facing the geometric center of the transmissive RIS array on the other side with distances from 50 \text{cm} to 500 \text{cm} with an interval of 50 \text{cm}, namely $r_r = 0.5 \sim 5\;  \text{m},\;\theta_{r} = 0^{\circ},\;\varphi_r = 0^{\circ}$. The results are illustrated in Fig.  \ref{fig:plot_dis}. It is easy to observe that the curves of the proposed model are close to that of the measurements. Specifically, the path loss increases with the increment of distances, and the growth rate is approximately quadratic. 

\begin{figure}[!htbp]
	\centering
	\includegraphics[width=0.33\textwidth]{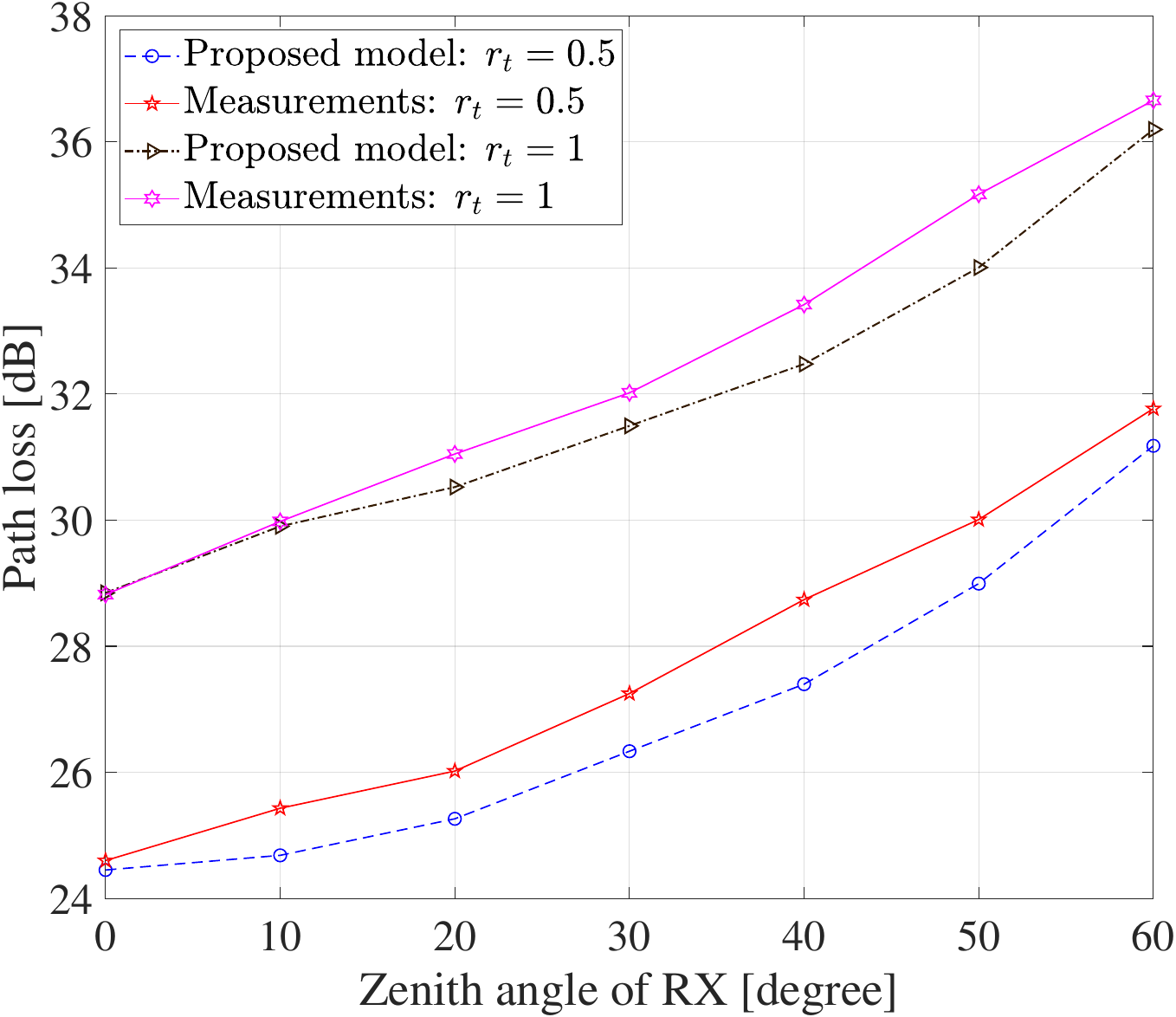}
	\caption{Path loss versus the zenith angle of RX $\varphi_r$ with $r_r = 4.5, \theta_{t} = 0, \theta_{r} = 0, \varphi_t = 0$.}
	\label{fig:plot_ang}
\end{figure}

Then, we study the impact of angles on the path loss. The transmitting horn antenna was placed in the same positions as the previous experiments, namely case 1: $r_t = 0.5\;  \text{m},\;\theta_{t} = \varphi_t = 0^{\circ}$ and case 2: $r_t = 1\;  \text{m},\;\theta_{t} =\;\varphi_t = 0$. Besides, the receiving horn antenna was placed at a fixed distance from the transmissive RIS array, as shown in Fig. \ref{fig:plot_ang}. Nevertheless, by adjusting the mechanical platform, we can change the zenith angles of RX from $0^{\circ}$ to $60^{\circ}$ with an interval of $10^{\circ}$, namely $\theta_{r} = 0^{\circ} \sim 60^{\circ},\;r_r = 4.5\; \text{m}, \;\varphi_r = 0^{\circ}$. Fig.  \ref{fig:plot_ang} illustrates the results of the path loss. We can observe that the simulation results are consistent with the measurements, which validates the accuracy of our proposed dual RCS-based path loss model. 

\subsection{Radiation Pattern Measurements}
In this section, we focus on the beam steering capabilities of the proposed transmissive RIS array and employ the fast optimization algorithm introduced in the previous section for measurements. The measurement was conducted under the same conditions as in section \ref{subsec:phase and amplitude measurement}. 

During the measurement process, we first ensured that the position of the receiving antenna remained fixed and then precisely rotated the central turntable in the microwave anechoic chamber. Specifically, the turntable was rotated from 0$^{\circ}$, in increments of 10$^{\circ}$, until reaching 60$^{\circ}$, covering a total of seven distinct angles. The purpose is to comprehensively assess the beam control performance of the transmissive RIS system at various angles. At each specified angle, the optimization algorithm ran once. 
After completing the optimization at each angle without altering the state of the array, we conducted detailed measurements of the radiation pattern using a vector network analyzer.

\begin{figure}[!htbp]
    \centering
    \includegraphics[width=0.43\textwidth]{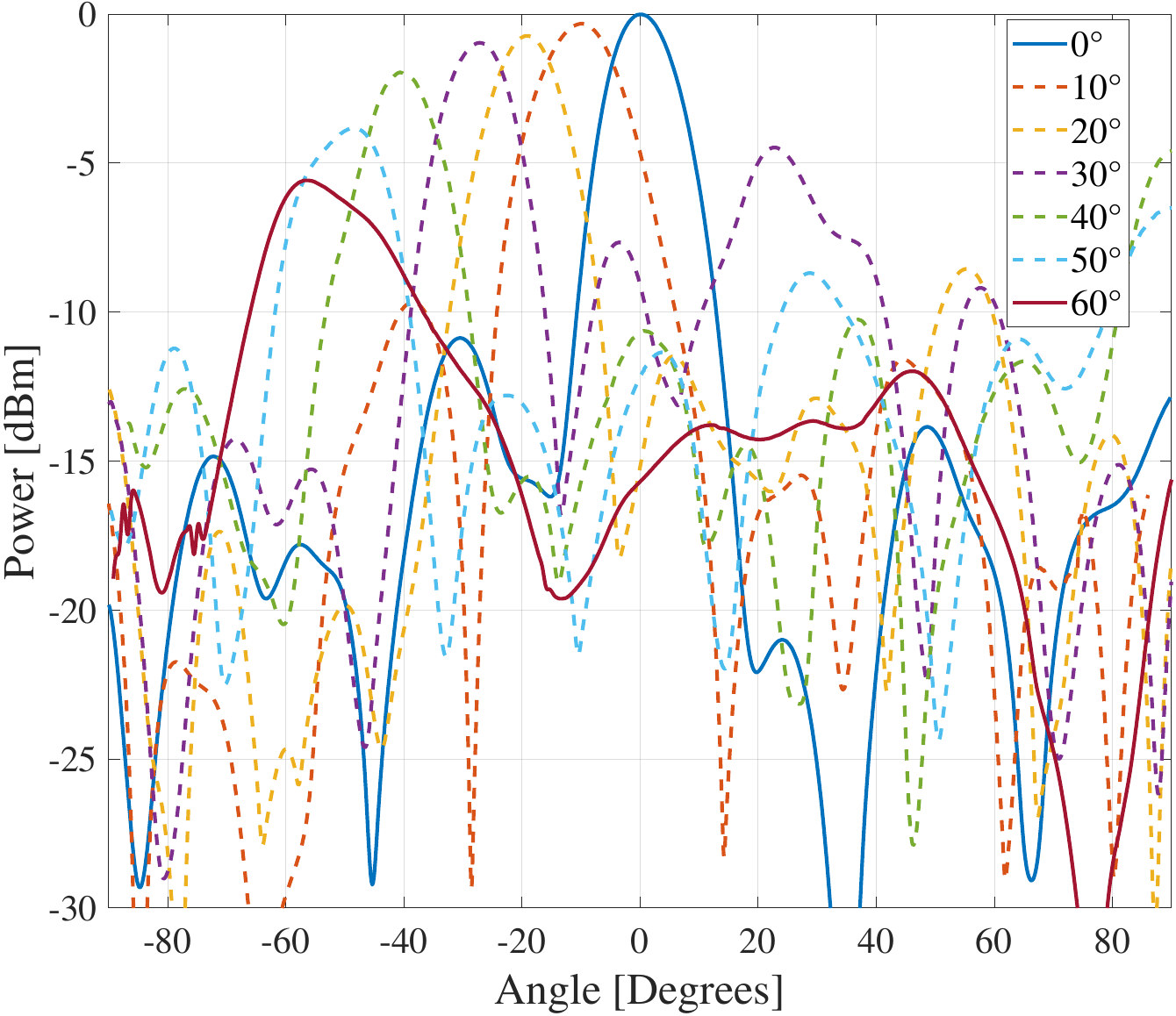}
    \caption{The measured radiation patterns of the RIS with pre-defined deflection angles.}
    \label{fig:radiation pattern}
\end{figure}

The directional patterns illustrated in Fig.~\ref{fig:radiation pattern} demonstrate the beam steering capabilities of the transmissive RIS system within a ±60$^{\circ}$ deflection range. The asymmetry in the radiation patterns can be attributed to factors such as incomplete structural symmetry of array elements, uneven element distribution, mutual coupling, physical effects from the testing environment, and manufacturing errors. Notably, the system exhibits best performance at a 0$^{\circ}$ deflection angle, where transmissive RIS directly aligns with the target direction. In this state, it achieves the highest gain and the narrowest main lobe width of 14$^{\circ}$. This combination of high gain and narrow main lobe width is an ideal response to precise directional transmission requirements.

As the beam deflection angle increases, a gradual decline in system performance is observed. When the deflection angle reaches specific values, such as ±50$^{\circ}$, the half-power beamwidth of the radiation pattern remains below 18$^{\circ}$, demonstrating good directional control capabilities, with only about a 3.5 \text{dB} reduction in gain performance. Furthermore, at a deflection angle of 60$^{\circ}$, the half-power beamwidth is observed to increase to 22$^{\circ}$. This phenomenon indicates that while transmissive RIS maintains specific beam steering capabilities over a wide angular range, the ability to direct and focus is reduced under conditions of extreme deflection angles.

Overall, measurement results confirm efficient performance in phase control, signal amplification, adaptability, and reliability in dynamic communication environments. The proposed dual RCS-based path loss model is validated in experiments. Additionally, the measurements underline the potential of this technology to enhance wireless communication systems and combat the double-fading effect of traditional RIS.

\section{Conclusion}

In this study, we proposed and validated an active transmissive RIS operating at the 2.6 GHz frequency band, which is used by China Mobile for 5G communications, to enhance signal transmission in various wireless communication scenarios. Our design demonstrates significant potential for applications such as enhanced cell-edge coverage, complex urban environments, and large indoor spaces. The active transmissive RIS significantly improves signal strength and quality by integrating power amplifiers and phase shifters, achieving up to 11.9 dB gain in signal power. Additionally, we introduced a preliminary full-duplex active transmissive RIS design based on FDD, enabling full-duplex capability in active RIS systems. We also introduced a dual RCS-based path loss model to accurately predict the received signal power in active transmissive RIS-aided wireless communication systems. Simulation and experimental results confirm the reliability and effectiveness of our design and model in practical scenarios. However, we acknowledge several limitations, including increased power consumption due to the integration of power amplifiers, relatively narrow bandwidth, and the need for further field trials in real-world communication environments. Despite these challenges, the active transmissive RIS offers substantial advantages over traditional passive RIS, particularly in terms of flexibility and signal penetration.

\ifCLASSOPTIONcaptionsoff
  \newpage
\fi

\bibliographystyle{IEEEtran}

\end{document}